\documentclass[12pt,preprint]{aastex}

\usepackage{graphicx}

\newcommand{\HI}{{\sc H\,i}}
\newcommand{\HII}{{\sc H\,ii}}
\newcommand{\kms}{\rm km\ s^{-1}}
\newcommand{\Tsys}{T_{\rm sys}}

\title{The VLA Galactic Plane Survey}  

\author{J. M. Stil\altaffilmark{1} and A. R. Taylor\altaffilmark{1}}
\author{J. M. Dickey\altaffilmark{2,3} and D. W. Kavars\altaffilmark{3}}
\author{P. G. Martin\altaffilmark{4,5}, T. A. Rothwell\altaffilmark{5}, and A. I. Boothroyd\altaffilmark{4}}
\author{Felix J. Lockman\altaffilmark{6}}
\author{N. M. McClure-Griffiths\altaffilmark{7}}

\altaffiltext{1}{Department of Physics and Astronomy, University of Calgary, 2500 University Drive NW, Calgary, AB, T2N 1N4, Canada}
\altaffiltext{2}{School of Mathematics and Physics - Private Bag 37, University 
of Tasmania, Hobart, TAS 7001, Australia}
\altaffiltext{3}{Department of Astronomy, University of Minnesota, 116 Church Street, SE, Minneapolis, MN 55455, USA}
\altaffiltext{4}{Canadian Institute for Theoretical Astrophysics, University of Toronto, 60 St. George Street, Toronto ON, M5S 3H8, Canada}
\altaffiltext{5}{Department of Astronomy and Astrophysics, University of Toronto, 60 St. George Street, Toronto ON, M5S 3H8, Canada}
\altaffiltext{6}{National Radio Astronomical Observatory, P.O. Box 2, Green Bank, West Virginia 24944, USA}
\altaffiltext{7}{Australia Telescope National Facility, CSIRO, P.O. Box 76, Epping NSW 1710, Australia}

\shorttitle{The VLA Galactic Plane Survey}
\shortauthors{Stil et~al.}

\begin{document}

\begin{abstract}
The VLA Galactic Plane Survey (VGPS) is a survey of \HI\ and 21-cm
continuum emission in the Galactic plane between longitude $18\degr$
and $67\degr$ with latitude coverage from $|b| < 1\fdg3$ to $|b| <
2\fdg3$. The survey area was observed with the Very Large Array (VLA)
in 990 pointings.  Short-spacing information for the \HI\ line
emission was obtained by additional observations with the Green Bank
Telescope (GBT).  \HI\ spectral line images are presented with a
resolution of $1\arcmin \times 1\arcmin \times 1.56\ \kms$ (FWHM) and
rms noise of $2\ \rm K$ per $0.824\ \kms$ channel. Continuum images
made from channels without \HI\ line emission have $1\arcmin$ (FWHM)
resolution.  The VGPS images reveal structures of atomic hydrogen and
21-cm continuum as large as several degrees with unprecedented
resolution in this part of the Galaxy.  With the completion of the
VGPS, it is now possible for the first time to assess the consistency
between arcminute-resolution surveys of Galactic \HI\ emission. VGPS
images are compared with images from the Canadian Galactic Plane
Survey (CGPS) and the Southern Galactic Plane Survey (SGPS).  In
general, the agreement between these surveys is impressive,
considering the differences in instrumentation and image processing
techniques used for each survey. The differences between VGPS and CGPS
images are small, $\lesssim 6\ \rm K$ (rms) in channels where the mean
\HI\ brightness temperature in the field exceeds $80\ \rm K$. A
similar degree of consistency is found between the VGPS and SGPS.  The
agreement we find between arcminute resolution surveys of the Galactic
plane is a crucial step towards combining these surveys into a single
uniform dataset which covers 90\% of the Galactic disk: the
International Galactic Plane Survey (IGPS). The VGPS data will be made
available on the World Wide Web through the Canadian Astronomy Data
Centre (CADC).

\end{abstract}



\section{Introduction}

The physical processes in the feedback cycle of matter between stars
and the interstellar medium play an important role in the evolution of
galaxies, and in the way products of stellar nucleosynthesis are
dispersed. A quantitative description of these processes requires
knowledge of the poorly known topology of different phases of the
interstellar medium and the timescales and locations of transitions
between these phases. Atomic hydrogen is the link between the gas
heated or expelled by massive stars and cold molecular gas from which
new stars form. \HI\ is widely distributed and, within certain limits,
easily observable across the Galaxy through the 21-cm line.  As such,
atomic hydrogen has been used to study the dynamics of the Galaxy and
physical conditions in the diffuse interstellar medium.

To study the interstellar medium in transition, from the cold phase to
the warm phase or vice versa, or the dynamical effect of stars on the
interstellar medium, requires observations which reveal parsec-scale
structures.  For objects outside the solar neighborhood, resolving
parsec-scale structures requires arcminute-resolution images. The
distribution of cold atomic gas may be revealed by absorption of
continuum emission or absorption of the \HI\ line emission itself
(\HI\ self absorption). Results from such observations depend strongly
on the resolution of the data \citep{DL1990}.  To place these
processes into a Galactic context, the high resolution image should
also show structure on large scales. At 21 cm wavelength, images with
arcminute resolution and a large spatial dynamic range can be obtained
by an interferometer in combination with a large single dish telescope
to fill in large scale structure not detected by the interferometer.

Previously the Canadian Galactic Plane Survey (CGPS)
\citep{taylor2003} and the Southern Galactic Plane Survey
\citep{mcclure2001,mcclure2005} have provided high-resolution \HI\
images in the northern sky (mainly the second Galactic quadrant) and
the southern sky (third and fourth Galactic quadrants). The CGPS will
be extended to longitude $50\degr$, while the SGPS has been extended to
longitude $20\degr$. A large part of the first Galactic quadrant is
located near the celestial equator. This area of the sky cannot be
observed at sufficient angular resolution by the interferometers used
in the CGPS and SGPS, because these interferometers have exclusively
or mostly east-west baselines. At the extremes of the CGPS and SGPS,
the angular resolution of these surveys is degraded by a factor $\sim
3$ in declination.

The Very Large Array (VLA) can observe the equatorial part of the
Galactic plane with adequate resolution. In this paper we present the
VLA Galactic Plane Survey (VGPS). The VGPS is an \HI\ spectral line
and 21-cm continuum survey of a large part of the first Galactic
quadrant. This survey was done with the Very Large Array (VLA) and the
100 m Robert C. Byrd Green Bank Telescope (GBT) of the National Radio
Astronomical Observatory (NRAO).  Short spacing information for the
VGPS continuum images was provided by a continuum survey
with the 100 m Effelsberg telescope \citep{reich1986,reich1990}. The
CGPS, SGPS, and VGPS will be combined into a single data set which
provides arcminute-resolution images of Galactic \HI\ for 90\% of the
Galactic plane as part of the International Galactic Plane Survey
(IGPS).  With the completion of the VGPS, different parts of the IGPS
overlap for the first time, allowing a detailed comparison of the
results from each survey. This paper describes the VGPS data, and
compares the VGPS spectral line images with those of the CGPS and SGPS
in the areas where the surveys overlap.

\section{Observations and data reduction}

\subsection{VLA observations}
\label{observe-sec}

The main set of observations of this survey was done with the Very
Large Array (VLA) of the NRAO.  The technical specifications of the
array are described in detail by \citet{TUP2003}.  The VLA is an
interferometer with 27 elements, each 25 m in diameter ($32'$ FWHM
primary beam size at 21 cm). Several VLA fields must be combined in a
mosaic to image an appreciable part of the Galaxy.  The most compact
configuration of VLA antennas, the D-configuration, has baselines
between 35 m and 1.03 km, and is the most suitable for imaging of
widespread Galactic \HI\ emission. For short observations (snapshots),
the largest angular scale that can be observed reasonably well is
$\sim$450 arcseconds at 21 cm.  The synthesized beam size is about 45
arcseconds (FWHM) at 21 cm. The total amount of observing time at the
VLA allocated to this survey was 260 hours in the period July to
September 2000. In addition to the VLA observations, a fast survey
with the 100 m GBT of the NRAO was done to obtain the necessary
short-spacing information for the \HI\ spectral line images.
Table~\ref{VGPSpar-tab} lists basic parameters of the VGPS.
  
\subsubsection{Mosaic Strategy}

An interferometer survey using the mosaicking technique to combine
many primary beam areas into a large map begins with the choice of the
area to cover and the telescope time available.  These two numbers set
the overall sensitivity or noise level of the survey, but this
sensitivity is not uniform over the area. The gain in a single VLA
field varies over the $32'$ (FWHM) primary beam of the VLA antennas.
When multiple fields are combined in a mosaic, the spacing between
pointing centers determines the corrugation or spatial variation of
the sensitivity function.  A centered hexagonal geometry is optimum
for flattening the sensitivity function, for a given number of
pointings, but the function can always be flattened further by
decreasing the spacing between adjacent beam centers and so increasing
the total number of pointings observed.  The overhead costs in
telescope drive time and the complexity of the data reduction are
increased as the number of pointings increases, so there is a
compromise required between flattening the sensitivity function and
reducing the number of pointings.  For this survey we have used the
relatively wide spacing of $25\arcmin$ between points, which is not
much less than the full-width to half-power point of the VLA beam at
$\lambda$ 21-cm of $32\arcmin$.  This results in the sensitivity
function shown in Figure~\ref{jd1}.  The $25\arcmin$ spacing choice is
driven by the 20 second minimum lag time between the end of data
taking on one scan and the beginning of the next, which is exacerbated
by the need to delete the first 10 second data average of each scan.
These features of the current VLA data acquisition system make it a
relatively slow telescope for mosaicking.  There are other options
available for data taking while driving the telescope (mode OF), but
these proved to be impractical for this survey.

\begin{figure}
\resizebox{\textwidth}{!}{\includegraphics[angle=-90,clip]{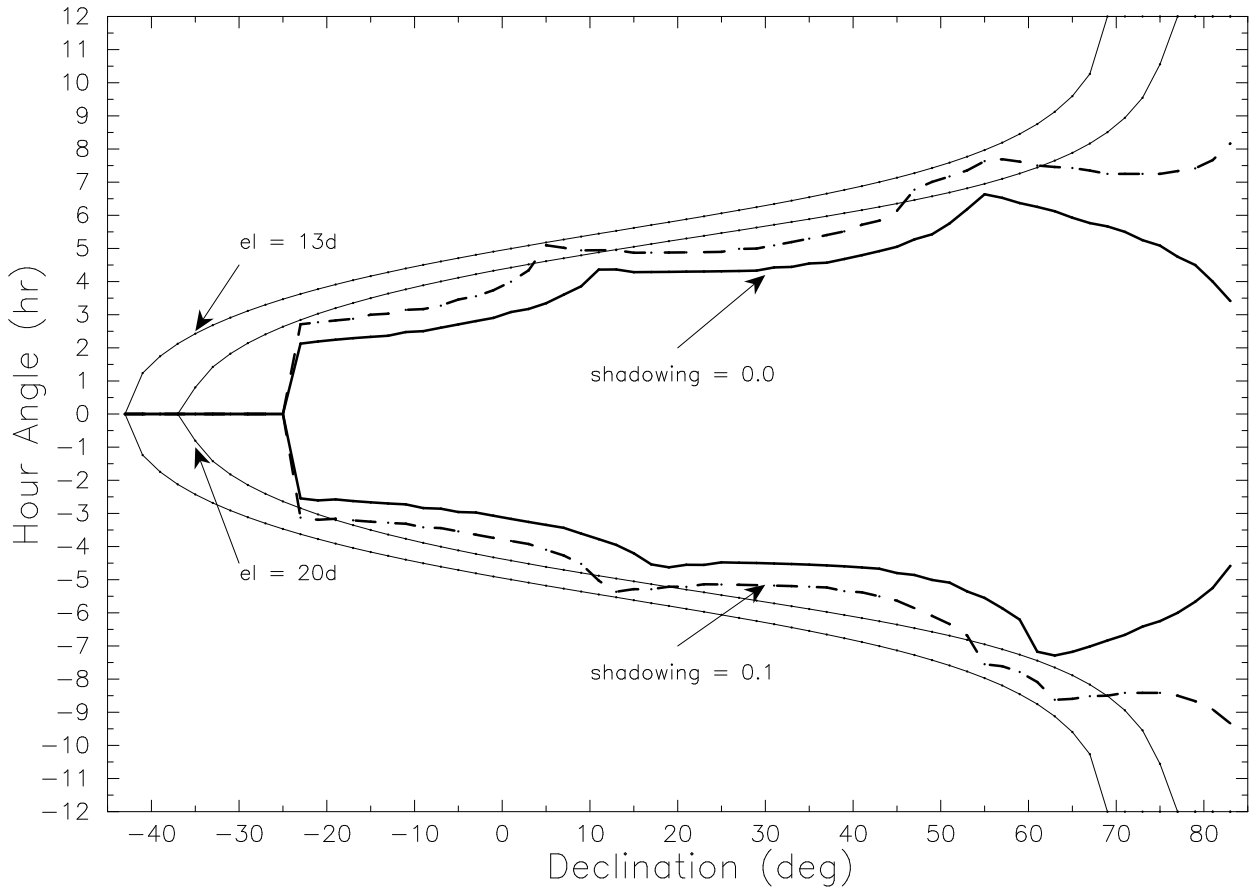}}
\caption{ Shadowing vs. hour angle and declination for the D array of
the VLA.  Two elevations, $13\degr$ and $20\degr$, are plotted with
thin lines; two shadowing limits, zero shadowing and 10\% shadowing,
are plotted with thick lines.  These limits were computed using the
AIPS subroutines GETANT, UVANT, and BLOCK, for the antenna file
corresponding to the array used for all survey observations.  All
observations of the primary survey area were taken with zero
shadowing, i.e., for hour angles in the area between the solid curves.
Some of the observations of the higher latitude fields (($|b| >
1\degr$) were taken with some shadowing, but never more than a few
percent.
\label{jd2}
}  
\end{figure}

The D configuration of the VLA is subject to shadowing, i.e. blockage
of one antenna by another, at low elevations.  For a mosaic survey, it
is even more important than usual to avoid shadowing, because of its
effect on the primary beam shape of the blocked antenna, which
compromises the estimate of the telescope response to a model source
that is needed for the maximum entropy deconvolution.  This survey
includes many fields at negative declinations, for which the available
hour angle range is minimal. Avoiding shadowing becomes the first
driver of the observing strategy.  Figure~\ref{jd2} shows a rough
guide for the hour angle and declination range which is safe from
shadowing for the D configuration.  All scans for the primary survey
area ($|b| < 1\degr$, $18\degr < l < 65\degr$) were taken without
shadowing, as were most in the secondary area ($|b| > 1\degr$).

\begin{figure}
\resizebox{\textwidth}{!}{\includegraphics[angle=-90,clip]{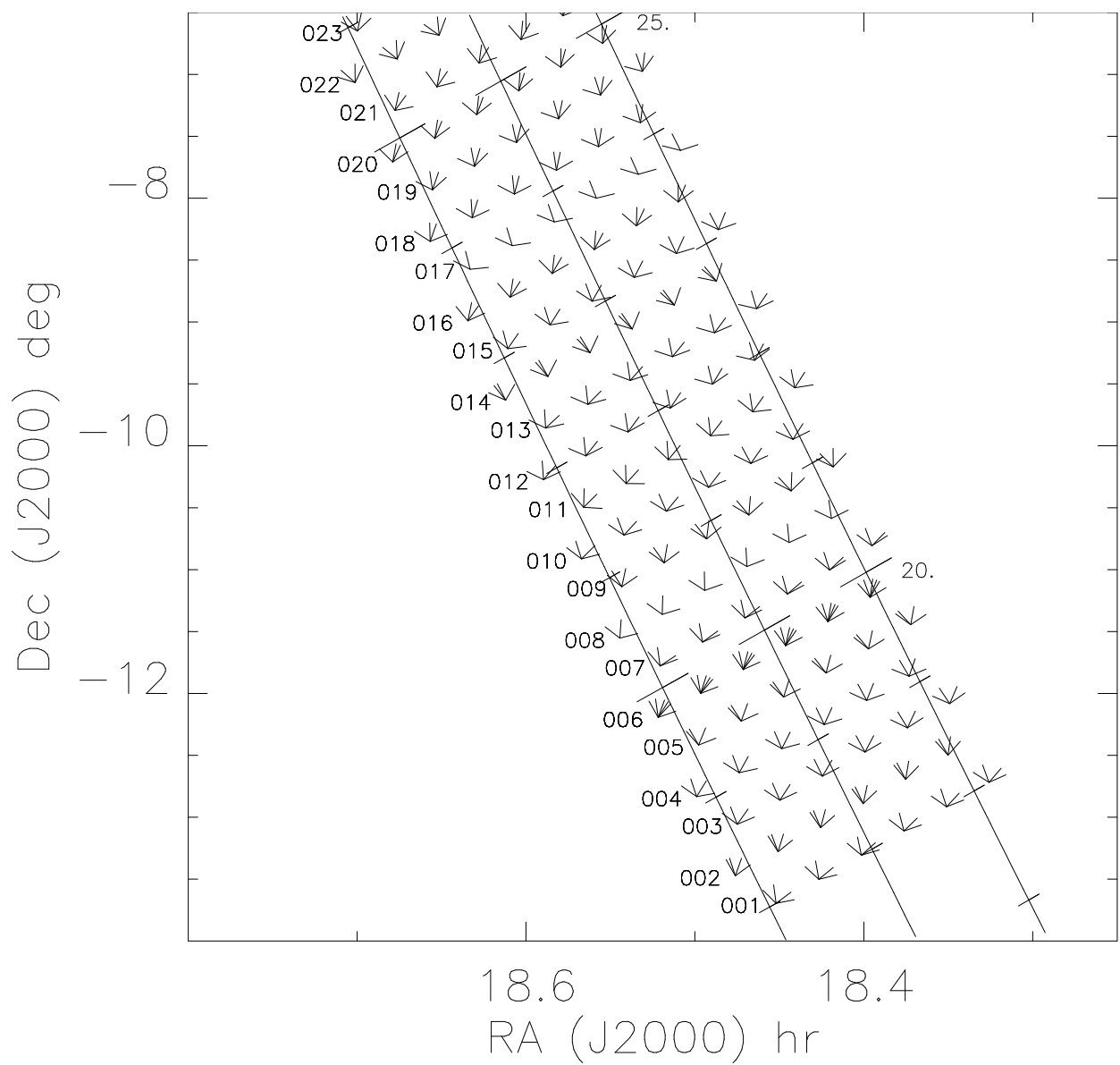}}
\caption{ Hour angles of the observations of sample pointings.  The
Galactic plane follows the center line, and the two lines on either
side show latitudes $-1\degr$ and $+1\degr$.  Longitudes are marked
with short bars, with longitude $20\degr$ and $25\degr$ indicated.
The pointing centers are shown as the vertices of several short line
segments (``chicken feet'').  The short segments each represent an
observation, with the direction of the segment indicating the
corresponding hour angle, with $0^{\rm h}$ plotted as a vertical
segment, $-6^{\rm h}$ and $+6^{\rm h}$ plotted horizontally to the
left and right.  The observing schedule was chosen to optimize the
{\it uv} coverage provided by a wide range of hour angles, i.e., a
broad footprint for each pointing.
\label{jd4}
}  
\end{figure}

Besides avoiding shadowing and minimizing telescope drive times, the
most important consideration for scheduling was obtaining multiple
scans on each field at widely spaced hour angles.  This is the best
way to minimize sidelobes due to large unsampled regions of the {\it
uv} plane that compromise the dynamic range of the resulting maps.
The scheduling process was driven by the need to spread the
observations of each field over the available hour angle range at the
declination of that field, and still fit everything into about 25
sessions, each typically covering $14^{\rm h} < {\rm LST} < 24^{\rm
h}$.  The low longitude end, which is observable only from $\sim
16^{\rm h}$ to $\sim 20^{\rm h}$ LST, was hard to fit into
approximately 25 sessions.  The area north of $+10\degr$ declination
($l > 45\degr$) is relatively easy to schedule. For simplification,
each row of pointings at constant longitude (six beam areas, 20
minutes of integration time total) was scheduled in a block.  The
final observing sequence gave hour angle coverages as shown in
Figure~\ref{jd4}.  Top priority for scheduling was given to the
primary area of the survey, latitude $-1\degr$ to $+1\degr$, longitude
$18\degr$ to $65\degr$.  The fields at higher positive and negative
latitudes were observed with second priority, so their {\it uv}
coverage is less evenly distributed over the available hour angle
range.  Generally the strongest continuum sources that cause the worst
dynamic range problems are located in the primary area, so this
strategy is appropriate.  However it should be kept in mind that the
quality of the imaging in the area of the survey with $|b|>1\degr$ is
degraded relative to the lower latitudes.  Most fields (93\%) were
observed at least three times at different hour angles, and 38\% of
the fields were observed four or more times.  About 7\% of the fields
were observed only 2 times. The theoretical sensitivity of the
spectral line mosaics is 8.4~mJy (1 $\sigma$) or 1.4~K per $0.824\
\kms$ channel for the beam size of $60\arcsec$.  The rms noise
amplitude in the final VGPS images is $1.8\ \rm K$ per channel on
average. The actual spatial variation of the noise in the final VGPS
images deviates somewhat from the regular pattern shown in
Figure~\ref{jd1} because of differences in the integration time per
field. Such differences exist because of variation in the number of
visits to a field and because of a longer dwell time on the first
field of a block of six.

\subsubsection{Spectrometer}

\begin{figure}
\resizebox{\textwidth}{!}{\includegraphics[angle=-90]{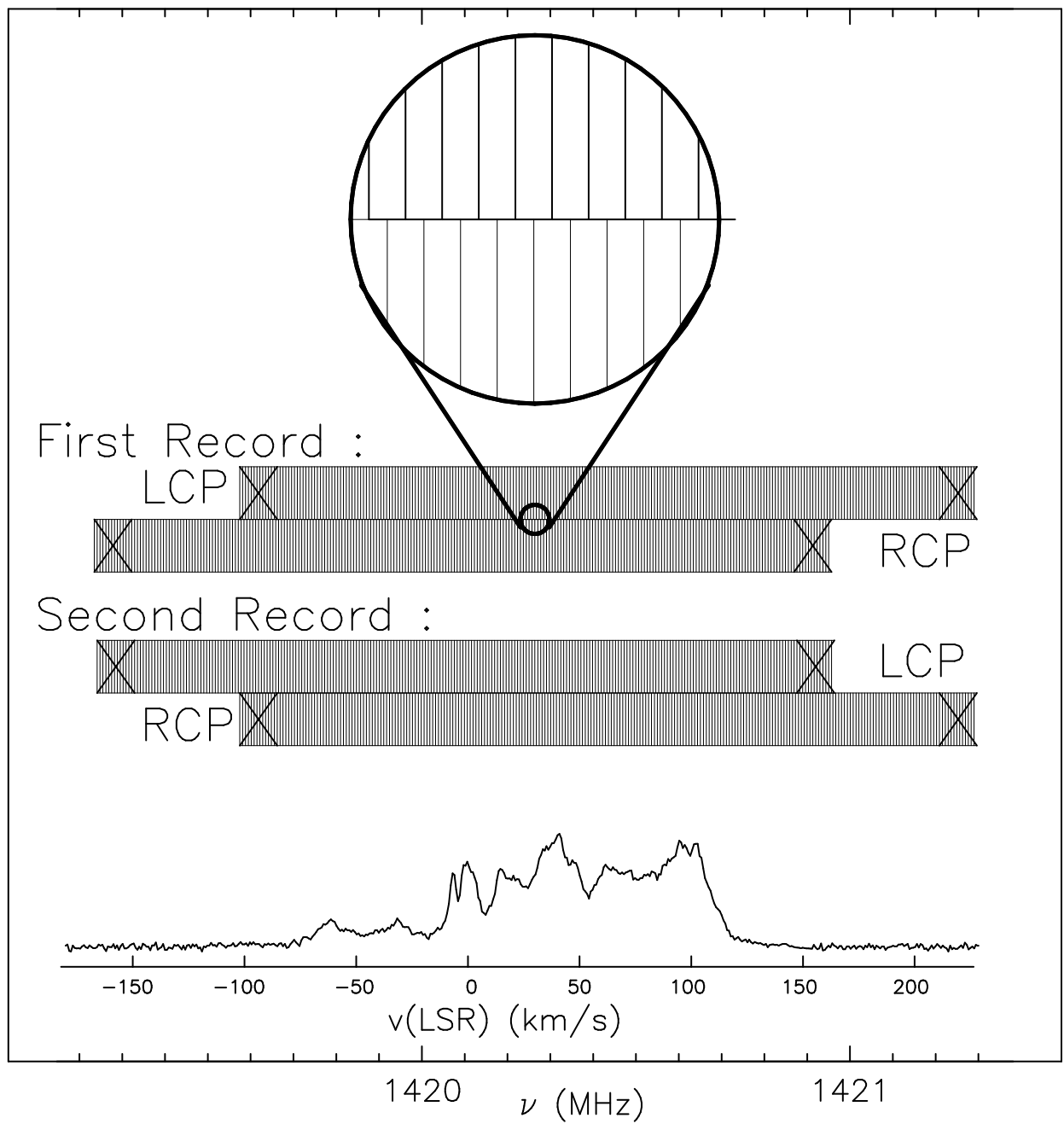}}
\caption{ Spectrometer channel spacing.  Observing frequency is
indicated on the bottom, with a representative spectrum shown in the
inset, with velocity shifted by a typical offset from the rest
frequency of 1420.4058 MHz due to terrestrial and solar motion.  All
observations were done in pairs, with the center frequencies of the
two polarizations switched as indicated.  The channel center
frequencies are staggered as shown in the magnified inset, so as to
make it possible to sample the profile shape with $0.824\ \kms$
channels in spite of the necessity to take the data with a broader
channel spacing ($1.28\ \kms$).  The center velocity was set at
$v_c(l)=+80-(1.6 \times l)$ for each longitude, $l$.
\label{jd5}
}  
\end{figure}

This survey pushes the limits of the VLA correlator in that we need
high resolution in velocity ($0.824\ \kms$ = 3.90 kHz channel spacing)
and broad bandwidth ($\sim 300\ \kms$ = 1.4~MHz total).  This was
impossible to obtain with the existing spectrometer for two
polarizations at once.  To sacrifice one of the two circular
polarizations would be equivalent to giving up half the integration
time of the survey, so we chose a strategy that keeps both
polarizations but with the coarser velocity channel spacing of $1.28\
\kms$.  We then stagger the placement of the channels between the two
polarizations by half a channel spacing, $0.64\ \kms$, so that the
sampling on the spectral axis may be increased when all data are
combined (Figure~\ref{jd5}). The VGPS spectral line data are sampled
on the same $0.824\ \kms$ spectral channels as the CGPS for maximum
consistency between the two datasets. The spectral resolution of the
data ($1.21 \times 1.28\ \kms = 1.56\ \kms$) is determined by the size
of the time lag window in the spectrometer. It is not changed by
resampling to narrower spectral channels.  The centers of the two
polarizations are offset by $+32.304\ \kms$ and $-31.460\ \kms$ from
the nominal center velocity, which is set at $v_c(l)=+80-(1.6 \times
l)\ \kms$ with $l$ the longitude in degrees.  Combining these gives
spectra with velocity width $341\ \kms$ after dropping 20 channels on
either side of the band due to the baseband filter shape.  The \HI\
line emission is unpolarized except for tiny amounts due to the Zeeman
effect, which are far below our sensitivity limit.  But to avoid
spurious spectral features arising from \HI\ absorption of linearly
polarized continuum, which is common in the synchrotron emission at
low latitudes, we alternate the frequency settings between the two
polarizations every 100 seconds.  So a single observation of a survey
field consists of two short integrations (100s each) with
complementary spectrometer settings.  This gives enough bandwidth to
cover the range of velocities in the first Galactic quadrant, +150 to
$-80\ \kms$ at the lower longitudes, with $0.824\ \kms$ channels
throughout.  The local oscillator settings for the survey were $-3.2$,
$3590$ with bandwidth code 5 (1.5625~MHz, of which the inner $\sim$
85\% is usable) and correlator mode 2AD.  No on-line Hanning smoothing
was performed, and the single dish bandpass shape was not used to
normalize the spectra, as is sometimes done as part of the correlation
step.

\subsection{VLA Calibration}
\label{calibration-sec}

Calibration of the VLA data was carried out using standard procedures
within AIPS. The primary calibrators 3C286 and 3C48 were used for flux
and bandpass calibration. After calibration, the visibility data were
imported into MIRIAD for further processing. Editing out glitches in
the large volume of visibility data for the 990 VLA fields was done
with an automated flagging routine.  The visibility data for the VGPS
fields were searched for high-amplitude points relative to the median
visibility amplitude for a particular baseline in a particular channel
and for a particular spectrometer/polarization combination. The
thresholds applied in this procedure were chosen after careful
inspection of the data.  Amplitudes more than 20 Jy above the median
amplitude in the duration of the snapshot were flagged.  Also, scans
with an overall median amplitude above 50 Jy were flagged to eliminate
saturated antennas. Special care was taken not to label legitimate
signal on the shortest baselines as bad data.  The snapshots are
sufficiently short that a constant visibility amplitude can be assumed
for each baseline in a single channel in this search.

The flagging procedure allowed streamlined visual inspection of
identified bad data and, if necessary, a human veto before the actual
flagging.  No false rejections were found because the rejection
criteria were sufficiently conservative.  If visibilities for a
particular combination of time, baseline and polarization were
rejected in one channel, all channels were rejected so as to keep the
{\it uv} sampling identical for all channels.  Lower amplitude
glitches often had higher amplitude counterparts in other
channels. The policy to flag all channels was found to be effective in
eliminating low-amplitude glitches as well. After the automated
flagging procedure, only a few incidental manual flagging operations
were required to make spectral line and continuum images free from
noticeable effects of bad data.

Preliminary continuum mosaics were constructed from visibility data
averaged over channels outside the velocity range of Galactic \HI\
line emission. These mosaics were analyzed with an automated source
extraction algorithm to compare the fluxes of compact continuum
sources with fluxes in the NVSS survey \citep{condon1998}.  Sources
were labeled as suspected variables and removed from consideration if
the ratio of the absolute value of the difference between the NVSS and
VGPS fluxes and the mean of these fluxes was larger than 10 times the
formal error.  This comparison between the NVSS and VGPS showed that
fluxes of compact sources in the VGPS were on average 30\% less than
fluxes listed in the NVSS. The origin of this discrepancy is not
understood. It is believed to be the result of the higher system
temperature in the VGPS observations, which is in part the result of
bright \HI\ emission in the Galactic plane. The VLA has an automatic
gain control (AGC) system that scales the signal with the system
temperature, but visibility amplitudes are corrected for this scaling.
First we discuss various factors that affect the system temperature in
the VGPS. Later we derive a correction to the flux scale of the VGPS
to make it consistent with the NVSS.

\subsubsection{Contributions to the system temperature}

Compared with the NVSS, there are two important enhanced contributions
to the system temperature.  One contribution is from bright Galactic
\HI\ and continuum emission, which can double the system temperature
averaged over the inner 75\% of the frequency band. The system
temperature changes across the spectral band because the brightness of
the \HI\ line changes with velocity.  The other contribution is from
spillover to the ground when a field is observed at low
elevation. Emission from the atmosphere also depends on elevation,
adding 2 to 4 K to the system temperature. The effect of spillover is
an order of magnitude larger than this.  The average system
temperature of the VLA antennas in the zenith is $\Tsys \approx 35\
\rm K$. The system temperature increases rapidly at low elevation, to
approximately $70\ \rm K$ at elevation $30\degr$.  VGPS observations
were made over a wide range of hour angles to obtain adequate sampling
in the {\it uv} plane.  As a result, fields were regularly observed
far from the meridian at low elevation. In contrast, when the NVSS
was made, its fields were observed close to the meridian in order to
minimize ground noise. This strategy is more suitable for a continuum
survey which targets compact sources.

The system temperature for each antenna of the VLA is recorded and
stored with the visibility data. However, for the present purpose it
is more convenient to adopt a different measure of $\Tsys$ derived
directly from the visibility data.  The scalar average amplitude of
the continuum-subtracted visibilities (abbreviated here as ampscalar)
gives a spectrum that is proportional to the rms visibility amplitude
in each sample; it is noise dominated (proportional to $\Tsys$).  It
takes into account the editing of data rejected by the automated
flagging routine, and it is not necessarily averaged in frequency as
is the system temperature recorded with the data. After continuum
subtraction, only the shortest baselines contain some correlated
signal because of the Galactic \HI\ line. When averaging the ampscalar
data over all antennas, this remaining signal has a negligible
effect. This was verified by comparing the result with ampscalar
values averaged over baselines longer than $1 k\lambda$.  The
ampscalar values are proportional to the recorded system temperature
averaged over the array.

We write the total system temperature as the sum of the receiver
temperature $T_{\rm rec} \approx 35\ \rm K$, an elevation-dependent
term $T_{\rm earth}(h)$ which includes atmospheric emission but is
usually dominated by spillover to the ground, and the brightness
temperature of cosmic radio emission $T_b(v)$ which depends on
velocity because of the bright Galactic \HI\ line,
$$
\Tsys(v,h) = T_b(v) + T_{\rm earth}(h) + T_{\rm rec}   \eqno (1)
$$ The spectral line data of the VGPS allows separation of the
contribution of Galactic emission to $\Tsys$ from other
contributions. This in turn allows us to derive a new functional form
for the elevation dependence of $T_{\rm earth}(h)$.

\begin{figure}
\resizebox{\textwidth}{!}{\includegraphics[angle=0]{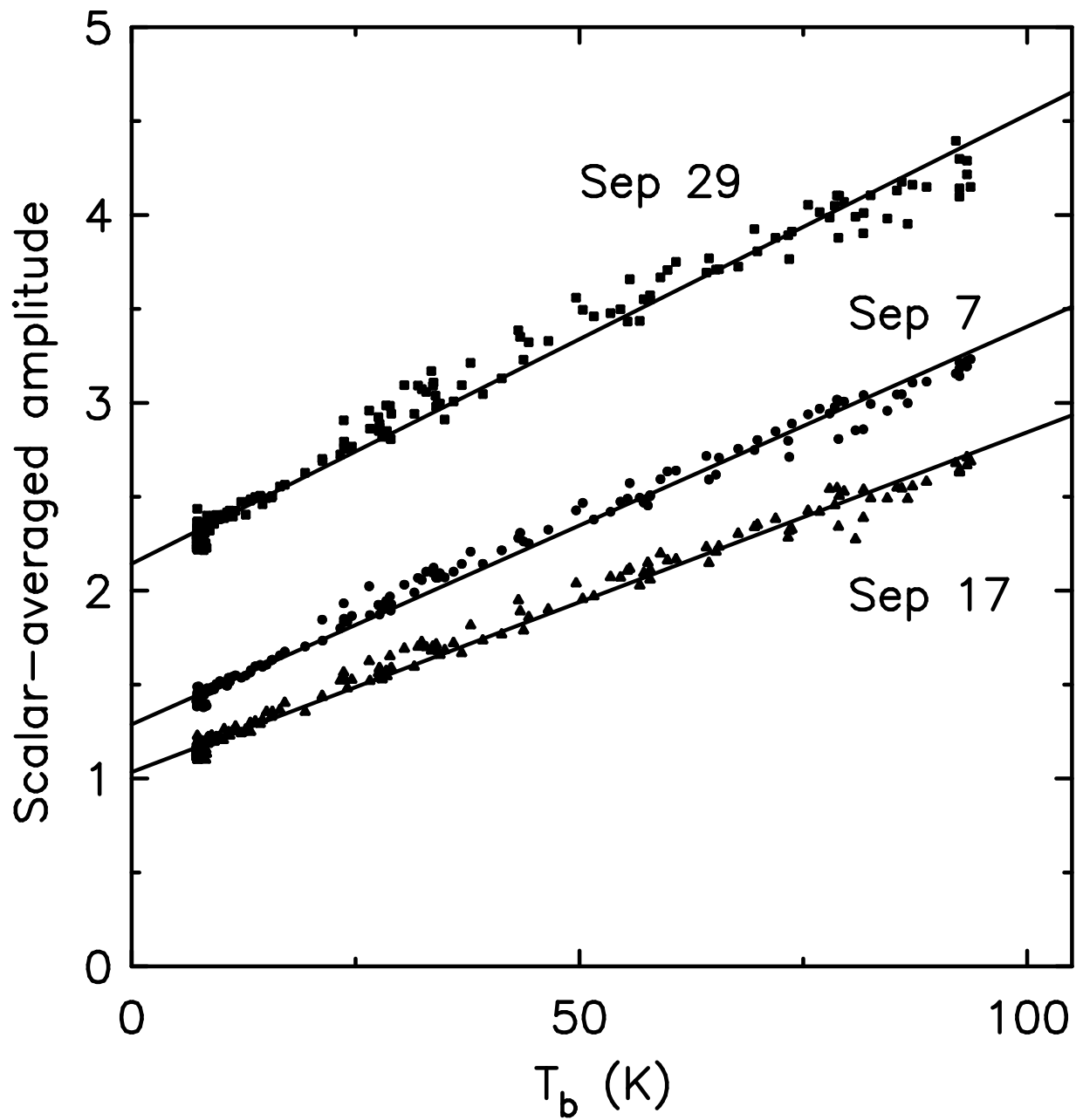}}
\caption{ Relation between scalar averaged visibility amplitude and
sky brightness temperature (line + continuum averaged over the VLA
primary beam) for three snapshots of the field G~65.3+1.4 taken with
the same spectrometer settings in a single polarization (L) on
2000 September 7, 17, and 29 at elevations $52\fdg3$, $63\fdg5$, and
$24\fdg2$ respectively. Each point corresponds to a single frequency
channel. The noise per channel changes as the brightness of the \HI\
line varies with velocity, and a significant variation of the noise
level with elevation is apparent.  The lines represent linear least
squares fits used to separate the contribution of Galactic radio
emission from other factors contributing to the noise.
\label{fitGBTVLA}
}  
\end{figure}

The brightness temperature in each velocity channel, $T_b(v)$, averaged
over the VLA primary beam, was determined by smoothing the GBT maps
and the Effelsberg maps to the resolution of the VLA primary beam.
Figure~\ref{fitGBTVLA} shows the relation between sky brightness
temperature (line + continuum) averaged over the primary beam of the
VLA, with the scalar averaged amplitude per channel.  Three visits to
the same field on three different days are shown.  The main difference
between the snapshots in Figure~\ref{fitGBTVLA} is the elevation of
the field at the time of observation. Some fields were observed at
nearly the same elevation on different days. Such observations have
nearly indistinguishable values of ampscalar. The relation between
ampscalar and brightness temperature was fitted with a linear relation
to allow extrapolation to $T_{b} = 0\ \rm K$. The contribution of
Galactic emission to the system temperature is eliminated by this
extrapolation.  We refer to this extrapolation as the scalar-averaged
amplitude at $T_b = 0$ or ampscalar at $T_b = 0$, which is
proportional to $T_{\rm earth}(h) + T_{\rm rec}$ according to Equation
(1).  We find that the slope of the relation in Figure~\ref{fitGBTVLA}
increases as the ampscalar at $T_b = 0$ increases.

\begin{figure}
\resizebox{\textwidth}{!}{\includegraphics[angle=-90]{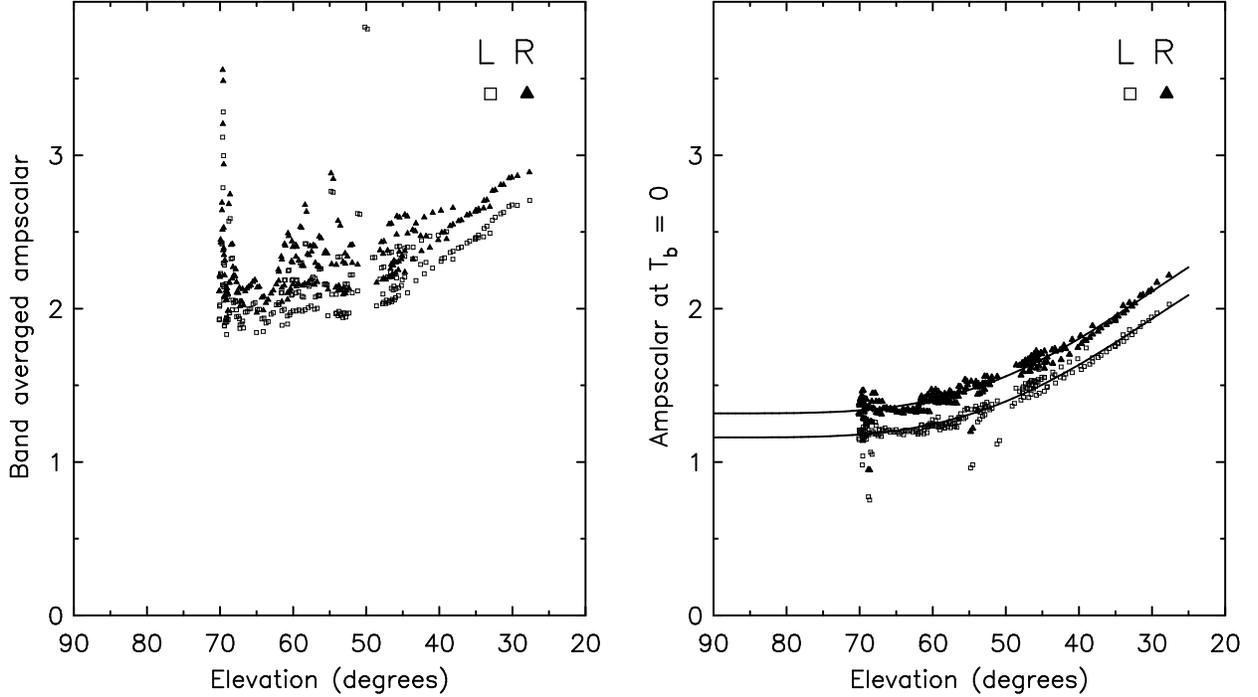}}
\caption{ Raw band averaged ampscalar (left panel) and ampscalar at
$T_b = 0$ (right panel) as a function of elevation for snapshots taken
on 2000 September 15.  This figure illustrates the magnitude of
contributions to the noise from Galactic \HI\ and continuum emission,
and spillover to the ground.  Left hand polarization is shown as open
squares, right hand polarization as filled triangles.  The bright
continuum sources W49 and W51 were observed on this day at elevations
$50\degr$ and $69\degr$ respectively. The curves in the right panel
represent fits of the form $A = a \cos^4(h) + b$, with $h$ the
elevation of the field.
\label{elevation}
}  
\end{figure}

Figure~\ref{elevation} shows the relation between the scalar-averaged
visibility amplitude and elevation of the field at the time of
observation for all snapshots taken on 2000 September 15. A poor
correlation is found between the raw band-averaged ampscalar and
elevation.  After correction for the contribution by Galactic
emission, a very tight relation is found between the ampscalar at $T_b
= 0$ and elevation. The scatter in this relation is consistent with
the estimated errors in the extrapolation to $T_b = 0$.  This
extrapolation is less accurate towards the brightest continuum sources
because the relation as shown in Figure~\ref{fitGBTVLA} is not well
defined. The points which do not fit on the relation represent
snapshots with very bright continuum.  Similar results were obtained
for each day of VGPS observations.

Figure~\ref{elevation} illustrates the relative importance of factors
which raise the system temperature. At high elevations, Galactic
emission roughly doubles the system temperature in the Galactic
plane. This increase is mostly due to the bright \HI\ line, but the
continuum also contributes 5 to 20 K to the system temperature, depending
on longitude. The brightest continuum sources (W49, and W51) actually
contribute more to the system temperature than the \HI\ line when
averaged over the spectral band.

The tight correlation between ampscalar at $T_b = 0$ and elevation
resembles the increase in $\Tsys$ with elevation from spillover to the
ground measured by \citet{TUP2003}, who applied a second-order
polynomial fit to the data. It is difficult to interpret the results
of a polynomial fit when comparing different days of observation,
because some days covered more fields at low elevation than others.
Inspection of all the data for each day separately showed that the
elevation dependence of ampscalar at $T_b = 0$ is also described
well with two free parameters by the form
$$
A = a \cos^4h + b  \eqno (2)
$$ 
Equation (2) was fitted to each day and polarization separately.
Three fields centered on ($l$,$b$) = ($43.3$,$-0.1$), ($49.0$,$-0.5$),
and ($49.4$,$-0.3$) were excluded from these fits. These fields are
most affected by W49 and W51. Results of the fits are listed in
Table~\ref{elevationfit-tab}.

Equation (2) provides an accurate fit of the elevation-dependent noise
contribution for each day.  Note that the scatter around the fits in
Figure~\ref{elevation} is much smaller than the difference between the
L and R polarization. The mean difference between the two
polarizations is $\langle b_L - b_R \rangle = 0.13 \pm 0.05$.  The
difference in ampscalar between the polarizations is equivalent to a
system temperature that is approximately 12\% higher in L than in R.
The rms residuals of the fits ($\sigma_L$ and $\sigma_R$ in
Table~\ref{elevationfit-tab}) are 0.06 in the mean. Inspection of the
fits indicated that larger values of $\sigma_L$ and $\sigma_R$
indicate variation in the system temperature during the day. Such
variation may be related to solar activity, in particular in September
when the angular distance of the Sun was smaller. Inspection of the
fits showed that variations in $a_L$ and $a_R$ in
Table~\ref{elevationfit-tab} appear to be related mainly to intra-day
variation. However, the range of values of $b_L$ and $b_R$ represents
real variations in the receiver temperature $T_{rec}$ during the
period of observations.

The fits in Table~\ref{elevationfit-tab} allow us to make an elevation
correction for the noise amplitude for each snapshot to obtain a
prediction of the noise level if the field had been observed at the
zenith. Figure~\ref{Ecor} shows the distribution per VGPS field of the
smallest value of the band averaged ampscalar of all snapshots
contributing to a field.  This map typically shows the ampscalar for
the observation of each field at the highest elevation. Most of the
structure in the distribution of the band averaged ampscalar
corresponds to the VGPS observing sequences of six fields in a row. A
few bright continuum sources near the Galactic plane can be identified
as well. The observing pattern visible in the distribution of the
band-averaged ampscalar shows that almost everywhere in the survey
area elevation-dependent contributions to $\Tsys$ are
important. Elevation-dependent effects and sky emission contribute
roughly equal amounts to the total system temperature
(Figure~\ref{elevation}).

The minimum ampscalar per field tends to be higher at low longitudes,
in part because these fields transit the meridian at a lower
elevation. Equation (2) allows a correction to be made for the
elevation of the field at the time of observation. The parameters $a$
and $b$ were determined for each day and for each polarization
separately by fitting Equation (2) to the available data as
illustrated in Figure~\ref{elevation}. The elevation-corrected
band-averaged ampscalar was calculated by subtracting the excess in
the band-averaged ampscalar value resulting from the elevation of the
field. The map of the elevation-corrected ampscalar in
Figure~\ref{Ecor} shows a higher noise level in the Galactic plane,
and toward bright continuum sources. Some residuals of the elevation
correction remain in this map. These are likely the result of
intra-day variations that are not traced by the daily average of the
elevation correction applied here.

The distribution of the elevation-corrected ampscalar shows a strong
resemblance to the VGPS continuum image. This resemblance is not
inconsistent with the fact that the \HI\ line usually contributes more
to the level of $\Tsys$ than the continuum. The visualization in
Figure~\ref{Ecor} tends to emphasize the variations between adjacent
fields.  Field-to-field variations in the velocity-integrated \HI\
brightness are fairly small compared with field-to-field variations in
the continuum level.

\subsubsection{Correction of the VGPS flux scale}

The remainder of this section describes a method to adjust the flux
scale of the VGPS to the flux scale of the NVSS.  The Canadian
Galactic Plane Survey also corrects its flux scale to the NVSS by
comparing the fluxes of continuum sources in each CGPS field as part
of the field registration process described in \citet{taylor2003}.  A
direct comparison of source fluxes per VGPS field is not possible
because, on average, only four compact continuum sources are available
per VGPS field.  Bright extended continuum emission from \HII\ regions
and supernova remnants also inhibits a comparison with NVSS sources.
A subset of the VGPS fields contains one or more isolated compact
continuum sources that can be compared with corresponding entries in
the NVSS catalog. We obtain an empirical relation between the NVSS to
VGPS flux ratio and the system temperature, including all
contributions, as measured by the band-averaged ampscalar. This
relation is used to predict a specific correction factor for each
snapshot.
 
To this end, each continuum snapshot was imaged and cleaned
individually.  Cleaning VLA snapshot images is difficult because of
the high level of sidelobes of the synthesized beam. This is a
particular concern for faint sources. Therefore, only sources with a
flux density larger than 100~mJy were considered. Sources identified
in the snapshot images were matched with sources in the NVSS
catalog. A source was accepted if its position was within $15\arcsec$
of the NVSS position, and if the deconvolved size of the source in the
NVSS catalog was less than $60\arcsec$. These requirements are aimed
to avoid misidentification while retaining a sufficient number of
sources to determine a relationship between the band averaged
ampscalar and the flux ratio NVSS/VGPS.
 
\begin{figure}
\resizebox{\textwidth}{!}{\includegraphics[angle=0]{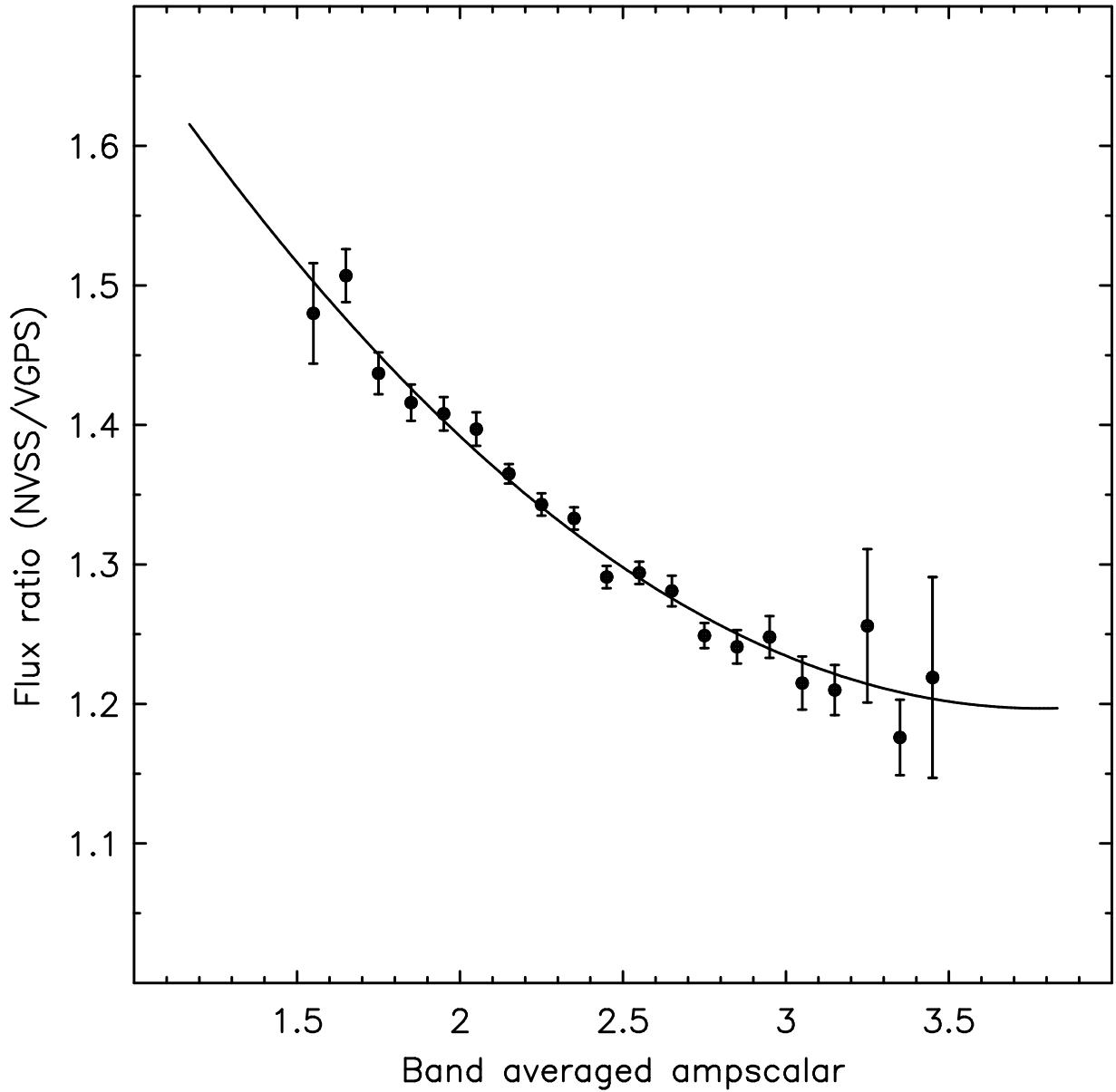}}
\caption{ The calibrating relation for the VGPS flux correction. The
flux ratio (NVSS/VGPS) is shown as a function of band averaged
ampscalar, which is proportional to $\Tsys$. The solid curve
represents the fit that is used to find the appropriate scale factor
for individual snapshots. After this correction, the fluxes of compact
continuum sources in all VGPS continuum mosaics were found to be
consistent with the NVSS to within 5\%. A detailed comparison of the
\HI\ spectral line images with the CGPS and SGPS (this paper) further
confirms the integrity of the flux scale after this correction. The
smaller correction for higher system temperatures is not understood.
\label{select_by_offset}
}  
\end{figure}

The snapshot images were sorted into narrow ranges of the band
averaged ampscalar. For each narrow range, a list of identified
sources was compiled.  Each of these source lists contains continuum
sources throughout the VGPS survey area detected in snapshots with
nearly the same band-averaged ampscalar. Figure~\ref{select_by_offset}
shows the ratio of NVSS flux to VGPS flux as a function of the band
averaged ampscalar. A tight relation is found between the band
averaged ampscalar and the flux ratio NVSS/VGPS. This relation defines
the correction factor for the flux scale of each snapshot based on its
band averaged ampscalar.  The scatter in the relation shown in
Figure~\ref{select_by_offset} is dominated by source number
statistics. The effect of the sample size was investigated by
regenerating the relation for fifty randomly selected subsamples, each
half the size of the total number of snapshots. The error bars in
Figure~\ref{select_by_offset} show the rms variation for each bin over
the fifty subsamples.  Values of the band-averaged ampscalar are
always larger than 1.4 in the VGPS.  A second-order polynomial was
fitted to the data to provide a prescription for the correction
factor, as a first-order polynomial fit was deemed insufficient.  A
small number of fields that include very strong continuum emission
have a band averaged ampscalar outside the range where the relation is
defined. We assume the relation to be constant for band averaged
ampscalar more than 3.8. All VGPS continuum mosaics were made again
with this correction to the flux scale. These new mosaics were
searched for compact continuum sources and the fluxes of these sources
were compared with fluxes listed in the NVSS catalog. This comparison
included many faint sources that were not included in the derivation
of the flux correction.  The flux scale of the VGPS mosaics was found
to be consistent with the NVSS flux scale within 5\%.

The success of applying the well-defined relation in
Figure~\ref{select_by_offset} adds confidence to the initial
assumption that the difference in the flux scale between VGPS and NVSS
continuum images was related to the higher system temperature in the
VGPS. However, the negative slope in this relation implies that the
largest correction to the flux scale is necessary for VGPS snapshots
with the lowest system temperature. This result remains unexplained.
However, we note that increased contributions to the system
temperature by bright continuum emission and by spillover to the
ground at smaller Galactic longitudes must also have occurred to some
extent in the NVSS observations.

The response of the VLA to increased system temperature has been
tested by following a flux calibrator during a time span of several
hours as it approaches the horizon. These experiments indicate that
the raw correlator output is adequately corrected for the scaling of
the signal by the AGC (R. A. Perley, private
communication). Calibrators observed for the VGPS are not well suited
to repeat this experiment because they cover a very limited range in
elevation. In particular, the primary flux calibrators were
consistently observed at high elevation. With limited significance, we
confirmed that the raw correlator output for the secondary calibrators
did not change with elevation. Our analysis shows that a calibrator
must be followed to very low elevations to probe the system
temperature regime of VGPS observations. The behavior of ampscalar at
$T_b=0$ in Figure~\ref{elevation} is representative for a calibrator,
because the $T_b(v)$ term in Equation (1) is very small for a
calibrator.  A comparison of the two panels of Figure~\ref{elevation}
shows that the system temperature in VGPS target fields almost always
exceeds the system temperature of a calibrator at an elevation of only
$25\degr$.  However, observations of a calibrator as it sets to
elevations as low as $10\degr$ \citep{TUP2003} cover the system
temperature range of most VGPS observations.

The relation in Figure~\ref{select_by_offset} compares flux
measurements of fairly bright, relatively isolated, compact continuum
sources. These flux measurements should not be affected by modest
differences in the {\it uv} coverage between the VGPS and the
NVSS. Since ampscalar values do not change if the minimum baseline is
taken as long as $1~ k\lambda$, the ampscalar values are also not
affected by differences in {\it uv} coverage. Therefore, we believe
that the behavior in Figure~\ref{select_by_offset} is not related to
differences in {\it uv} coverage between the VGPS and the NVSS.

In summary, the VGPS calibration follows these steps: 

\noindent
1. Standard gain and phase calibration was done in AIPS for all snapshots. 
The gain and phase corrections were applied to the data before importing
the visibilities into MIRIAD for imaging.

\noindent
2. Visibilities were channel-averaged (continuum images) or
continuum-subtracted (spectral line images) as appropriate. 

\noindent
3. The gain for each snapshot was adjusted by a single factor derived from its
band averaged ampscalar and the relation shown in Figure~\ref{select_by_offset}.
A small number of fields was selected for self-calibration after construction 
of the mosaics (Section~\ref{selfcal-sec}). 

\subsection{GBT observations}

Short-spacing information for the \HI\ line emission was obtained
using the Green Bank Telescope (GBT) of the National Radio Astronomy
Observatory (NRAO). The GBT is a 100 m dish with an off-axis feed arm
for an unblocked aperture which reduces the radio sidelobes, radio
frequency interference, spectral standing waves, and the effects of
stray radiation.  The extent of the GBT survey varies with longitude,
covering $|b| \le 1\fdg3$ for longitudes $18\degr\leq l \leq 45\degr$
and $|b| \le 2\fdg3$ for longitudes $45\degr \leq l \leq 67\degr$.
Observations of these regions began on 2002 November 21$-$25 and
continued on 2003 March 6$-$9, 2003 May 26$-$27, and 2003 August
29$-$30.  The observing strategy was to make small, $\Delta l =
2\degr-5\degr$, \HI\ maps `on the fly.'  Using this technique, the
telescope was driven at a rate of $3\degr$ per minute with a sample
written every second.  After driving through the full latitude range,
the telescope was stepped $3\arcmin$ in longitude.  This process was
continued until the particular map was completed.  The data were taken
in frequency-switching mode using the GBT spectral processor with a
total bandwidth of $5\ \rm MHz$ across 1024 channels.  The resulting
channel spacing is $1.03\ \kms$ and the spectral resolution is $1.25\
\kms$ (FWHM). Data were taken by in-band frequency switching yielding
a total velocity coverage of $530\ \kms$ centered at $+50\ \kms$ LSR.
IAU standard regions S6 and S8 \citep{williams1973} were observed and
used for absolute brightness temperature calibration.  The angular
resolution of the GBT data is $\sim 9\arcmin$.  The final GBT \HI\
spectra have an rms noise of $\sim 0.3\ \rm K$ in emission-free
channels.

Imaging and data calibration were done with the AIPS++ data
reduction package.  A first order polynomial was fitted to the
off-line channels to remove residual instrumental baseline structure.
Polynomial fits of higher order were also attempted, but these were
found to be no better than a first order polynomial fit.  Gridding
of the data was done with the task IMAGEMS, using a BOX gridding
function.  The latitude and longitude size of each cell was set to
$3\arcmin$, equal to the longitude spacing of the observations.  Each
of the small \HI\ maps had a narrow overlap in longitude with the
adjacent maps.  After applying the absolute brightness temperature
calibration, the overlap regions were compared for consistency.  A
majority of the overlap regions showed small discrepancies of order
1$-$3 K, but a few regions were found to have brightness temperature
differences of up to 6$-$8 K.  The overlap regions were used to scale
the cubes observed on different days to a common calibration, which is
consistent to within $3\%$.  As a second check of the observed
brightness temperature consistency, a small portion of the galactic
plane ($65\degr \leq l \leq 67\degr$; $-1.3\degr \leq b \leq
+1.3\degr$) was observed twice, once during the 2002 November 21$-$25
observing session and again during the 2003 August 29$-$30 session.
Only small differences of a few Kelvin were found between these maps.

\subsection{Imaging}

The definition of VGPS mosaic images is a direct extension of the set
of mosaic images of the CGPS. VGPS images are sampled in position
and velocity on the same grid as CGPS images. Mosaics of $1024 \times
1024$ pixels ($5\fdg12 \times 5\fdg12$) are centered at intervals of
$4\degr$ in longitude.  This provides significant overlap between
mosaics for coverage of objects near the mosaic boundary. The velocity
axis of VGPS spectral line cubes is sampled with the same channel
definitions as the spectral line cubes of the CGPS, but the velocity
range covered by the VGPS is different.  There are three VGPS data
products as described below: continuum images which include
short-spacing data, continuum-subtracted \HI\ spectral line images
which include short-spacing data, and continuum-included spectral line
images which do not include short-spacing data. All VGPS data products
will be made available on the World Wide Web through the Canadian
Astronomy Data Centre (CADC).

\subsubsection{Continuum}
\label{selfcal-sec}

VGPS line or continuum images are mosaics of several VLA pointings.
As it is impractical to process the entire survey area at the same
time, only fields with their central longitudes within $4\degr$ of the
center of the mosaic were included in the construction of each
mosaic. Each VLA field was imaged to the 10\% level of the primary
beam, and field images were combined with the VLA primary beam as
weighting function. The primary beam model for the VLA is the same as
used in the AIPS task LTESS.

Figure~\ref{snapshotUV} shows the sampling in the {\it uv} plane and
the synthesized beam for a representative field near the center of the
survey, composed of snapshots taken at three different hour angles. A
Gaussian weighting function in the {\it uv} plane was applied to
obtain a $60\arcsec$ (FWHM) synthesized beam throughout the survey
area. The strongest sidelobes of the synthesized beam are spokes
running radially outward from the main lobe in several position
angles. The pattern of the spokes depends on the hour angles at which
the field was observed. The pattern is different for each field, but
fields that were observed within the same sequence of six fields
(Section~\ref{observe-sec}), have similar but slightly rotated
patterns. The amplitude of the spokes in the dirty beam is typically
10\% of the main lobe, but peaks up to 17\% are usually present.  The
VGPS images must be deconvolved to remove artifacts resulting from
these sidelobes. Emission in the Galactic plane usually fills the
field of view. In this case, deconvolution is preferred after
mosaicking the fields.

Continuum images were constructed from visibility data averaged over
channels without discernible line emission, but avoiding noisy
channels near the edges of the frequency band.  Deconvolution of the
VGPS images was done with the MIRIAD program MOSMEM which uses the
maximum entropy method described by \citet{cornwell1985} and
\citet{sault1996}. Experiments showed that no significant improvement
was made after about 20 iterations of the algorithm, even if no formal
convergence could be reached. A maximum of 50 iterations was adopted
in the deconvolution of the continuum mosaics.  The criteria that
define the convergence of the deconvolution, the entropy function and
the $\chi^2$ criterium, include a summation over the entire image.
Therefore, the result of the deconvolution depends somewhat on the
area that was imaged in the construction of the mosaic. This causes
small differences between neighboring mosaics in the area where they
overlap, with rms amplitude at or below the level of the noise.

Standard calibration alone results in images with a maximal dynamic
range of $\sim$100, which is not sufficient to image bright continuum
sources in the survey area without discernible image artifacts.  The
continuum mosaics were inspected for residual sidelobes around bright
sources after the deconvolution step.  Self-calibration was attempted
for selected fields with strong artifacts, beginning with the field in
which the source is located closest to the field center.  A small
mosaic, typically including all fields within $\sim 1\degr$ of the
source, was made and deconvolved. If a particular visit to the field
could be identified as the prime origin of the artifacts, the
visibility data from this visit were excluded from the initial imaging
step. The deconvolution of this small mosaic produces a sky model of
``clean components'', which is used for phase self-calibration on the
central field only, after multiplication with the primary beam model
of the VLA. The small mosaic is then made with the improved
calibration solution for the central field.  If any data were left out
of the first imaging step, those data were included after the first
round of self calibration. If necessary, adjacent fields are
self-calibrated subsequently. The central field is usually
self-calibrated a second time after self-calibration of the
surrounding fields. After self-calibration, the final VLA continuum
mosaic was constructed and regridded to Galactic coordinates.

The results of the self-calibration are usually satisfactory. A
dynamic range of $\sim$200 was obtained in the continuum images after
self calibration on sources that mostly affect a single field.
However, self-calibration does not provide a good solution if the
source is too far from the field center. The brightest continuum
sources in the VGPS, in particular W49 and W51, generate artifacts
even at large distances from the field center. These artifacts remain
in the continuum images, mostly affecting fields surrounding
these sources. 

The VLA is not sensitive to emission on angular scales larger than
$\sim 30'$ because structures on these scales are resolved even by the
shortest projected baselines. The missing continuum short-spacing
information was provided by the continuum survey of \citet{reich1986}
and \citet{reich1990} with the 100 m Effelsberg telescope. The
deconvolved VLA mosaic was combined with the Effelsberg image with the
MIRIAD program IMMERGE. The interferometer and single-dish images are
both Fourier transformed and added in the {\it uv} plane with
baseline-dependent weights such that the sum represents the
visibilities for all angular scales weighted by a single Gaussian
weighting function defined by the $1'$ synthesized beam. More details
of this procedure were given by \citet{mcclure2005}. The single-dish
data may be multiplied with a factor close to 1 to compensate for a
potential difference in the flux calibration between the two
surveys. This factor can be determined by comparing the Fourier
transforms of the single-dish and interferometer images in a common
annulus in the {\it uv} plane, taking into account the difference in
resolution. This flux scale factor was found to be equal to 1 within
the uncertainties of a few percent, so it was set to 1 for all VGPS
continuum mosaics.

\subsubsection{\HI\ spectral line}

Continuum emission was subtracted in the {\it uv} plane by subtracting
the average of visibilities in the channels used to create the
continuum images.  The continuum-subtracted visibility data were
resampled from their original $1.28\ \kms$ channels to $0.824\ \kms$
channels before imaging with the velocity resampling method included
in MIRIAD. The resampled visibility data in an output channel are the
weighted mean of visibilities of all input channels that overlap with
the output channel. The weights are proportional to the amount of
overlap between each input velocity channel with the output velocity
channel.  VGPS velocity channels were defined to match with velocity
channels of the CGPS.  The resulting velocity sampling was tested by
comparing VGPS continuum absorption profiles with corresponding
profiles from other sources, in particular the CGPS. These tests
showed that the strategy of staggering velocity channels
(Figure~\ref{jd5}) and subsequent resampling of the data in MIRIAD was
successful in recovering a fully-sampled \HI\ line profile over the
required wide velocity range.  The resampled visibility data were
imaged and mosaicked in the same way as the continuum images.

Continuum sources in absorption appear as a negative imprint in the
\HI\ image in the form of the continuum source convolved with the
dirty beam. A continuum source brighter than $T_{\rm b} \approx 20\
\rm K$ displays visible sidelobes in channels where the optical depth
of the \HI\ line is significantly more than 1. The VGPS area contains
many brighter continuum sources which display significant sidelobes in
absorption, even if the optical depth of the \HI\ line is not that
high.  The maximum entropy deconvolution would consider the sidelobes
associated with absorbed continuum sources as structure in the \HI\
emission and try to deconvolve these structures. 

Before deconvolution of the \HI\ line emission, the sidelobes of
bright continuum sources are removed using the clean
\citep{hogbom1974} algorithm with the MIRIAD program MOSSDI.  A
temporary mosaic with (projected) baselines longer than $0.3\ \rm
k\lambda$ was made for this purpose. The exclusion of short baselines
in this mosaic eliminates most \HI\ emission and allows proper
cleaning of (negative) absorbed continuum sources without interference
from negative sidelobes associated with \HI\ emission.  The resulting
clean component model for absorbed sources was subtracted from the
\HI\ image and restored with a Gaussian beam before deconvolution of
the line emission. This procedure is effective in removing sidelobes
associated with absorbed continuum sources, in particular the radial
spokes that occur in the synthesized beam of VLA snapshots
(Figure~\ref{snapshotUV}).  The \HI\ line emission was then
deconvolved with the same maximum entropy algorithm as used for the
continuum images.  The deconvolved model of ``clean components'' of
the \HI\ emission was restored with a $1\arcmin$ (FWHM) circular
Gaussian beam. The synthesized beam in the VGPS \HI\ images is
therefore independent of position and identical to the synthesized
beam in the continuum images.

The clean \HI\ images were regridded to Galactic coordinates and
combined with the \HI\ zero spacing data set obtained with the
GBT. The process of combining the single-dish survey with the
interferometer survey allows rescaling of the single-dish data to
match the flux scale of the interferometer. The accuracy with which
this scale factor can be determined from the \HI\ spectral line data
was estimated at 5\% to 10\%. Within the uncertainties, no scaling of
the GBT flux scale was required. However, after a detailed comparison
of \HI\ images from the VGPS and the CGPS (see
Section~\ref{VGPSCGPS-sec}), it was decided to rescale the GBT data
with a factor 0.973 to obtain a single consistent calibration for the
CGPS and the VGPS. This correction factor is well within the
uncertainty of the absolute flux calibration. A consistent flux scale
for these surveys facilitates the comparison of properties of \HI\
emission in the inner and the outer Galaxy.

\subsubsection{Continuum-included \HI\ cubes}

\HI\ cubes were also constructed with both line and continuum emission
included, for use in 21-cm absorption line studies.  Instead of
subtracting the continuum (as was done for the continuum-subtracted
line cubes described in the previous section), the visibility data were
imaged directly with the same velocity channels as the spectral line
cubes. The weighting function in the {\it uv} plane was superuniform
weighting, which optimizes the angular resolution of the
images. Maximum angular resolution is more important for continuum
absorption experiments than brightness sensitivity.  A ``dirty''
continuum-included mosaic cube was created, and deconvolved using the
maximum entropy method; the deconvolved model of ``clean components''
was restored with a $50''$ circular Gaussian beam (slightly smaller
than the $60''$ Gaussian beam used to restore the line cubes and
continuum maps).  The deconvolved cubes were then regridded to
Galactic coordinates.  Note that {\em no zero-spacing data were
added\/} to these continuum-included H I cubes, since no zero-spacing
data set was available that included both line and continuum emission.

\section{Comparing the VGPS with CGPS and SGPS}

The three individual Galactic plane surveys will be combined into a
uniform high-fidelity multi-wavelength dataset which covers the
Galactic plane from $l = -107\degr$ to $l = +190\degr$ at least within
the latitude range $|b| < 1\degr$ with arcminute angular resolution.
This combined dataset is called the International Galactic Plane
Survey (IGPS). It will cover 90\% of the area of the Galactic disk.
With the completion of the VGPS, different IGPS components overlap in
narrow portions of the Galactic plane.  This allows for the first time
a comparison of these large mosaicking surveys of the \HI\ sky. The
success of an integrated IGPS dataset depends on the extent to which
the different surveys can produce consistent images when observing the
same part of the sky. This is a significant challenge given the
different telescopes and imaging techniques used, in particular for
the \HI\ line where extended bright emission is present throughout the
field of view.  Comparing the VGPS with the CGPS is of particular
interest because the image deconvolution step does not exist in the
CGPS.

\subsection{VGPS and CGPS}
\label{VGPSCGPS-sec}

The CGPS and the VGPS are each a combination of two separate surveys.
Low spatial frequencies are sampled by a single-dish survey, and high
spatial frequencies are sampled by an interferometer survey. The
single-dish survey dominates the signal at spatial frequencies smaller
than those sampled by the shortest projected baselines of the
interferometer. Most of the power in Galactic \HI\ emission occurs on
these large angular scales. The low-resolution survey for the CGPS was
done with the DRAO 26 meter telescope. The low-resolution survey for
the VGPS was done with the 100 m GBT.  These telescopes have different
resolution and stray radiation characteristics. Stray radiation
correction was done for the low-resolution DRAO survey
\citep{higgs2000}, but not for the GBT survey, as the sidelobes of the
GBT are much lower.

The shortest (unprojected) baseline of the DRAO synthesis telescope is
$60\lambda$, compared to $170\lambda$ for the VLA. The largest angular
scale sampled by the interferometer is therefore approximately 3 times
larger in the CGPS. On average, the VLA D-array also samples smaller
angular scales than the DRAO ST. The longest baselines of the VLA
D-array are $\sim 50\%$ longer than the longest baseline of the DRAO
ST. At the declination of the area of overlap ($\delta \lesssim
30\degr$) the array configurations and difference in latitude between
DRAO and the VLA further increase the difference between the
interferometers in projected baselines.

The most significant difference between the VGPS and the CGPS is the
density of the sampling of the aperture plane by the interferometer. A
full synthesis with the DRAO ST samples the aperture plane completely
between an inner and an outer boundary \citep{landecker2000}. This
complete sampling allows direct imaging of the visibility data without
deconvolution. The sampling of the aperture plane by VGPS snapshots is
less complete.  The necessary deconvolution of the VGPS images is a
priori the most likely source of differences between CGPS and VGPS
\HI\ images.

In order to compare the CGPS and VGPS spectral line data, a special
VGPS spectral line cube was made with a resolution that matches the
$122\arcsec\times 58\arcsec$ synthesized beam of the DRAO ST in the
overlap area. The major axis of the DRAO synthesized beam changes by
17\% across the area where the VGPS and CGPS overlap. Assuming an
average beam shape for the comparison leads to a maximum mismatch of
the beam size of less than 9\% in the declination direction in the
extreme corners of the area where VGPS and CGPS overlap. The spectral
resolution of the CGPS is $1.32\ \kms$ (FWHM), marginally smaller than
the $1.56\ \kms$ (FWHM) resolution of the VGPS. The difference is less
than half a frequency channel. This small difference in velocity
resolution is not expected to produce a noticeable difference because
of the modest dynamic range of the spectral line data.

\begin{figure}
\resizebox{\textwidth}{!}{\includegraphics[scale=.8,angle=0]{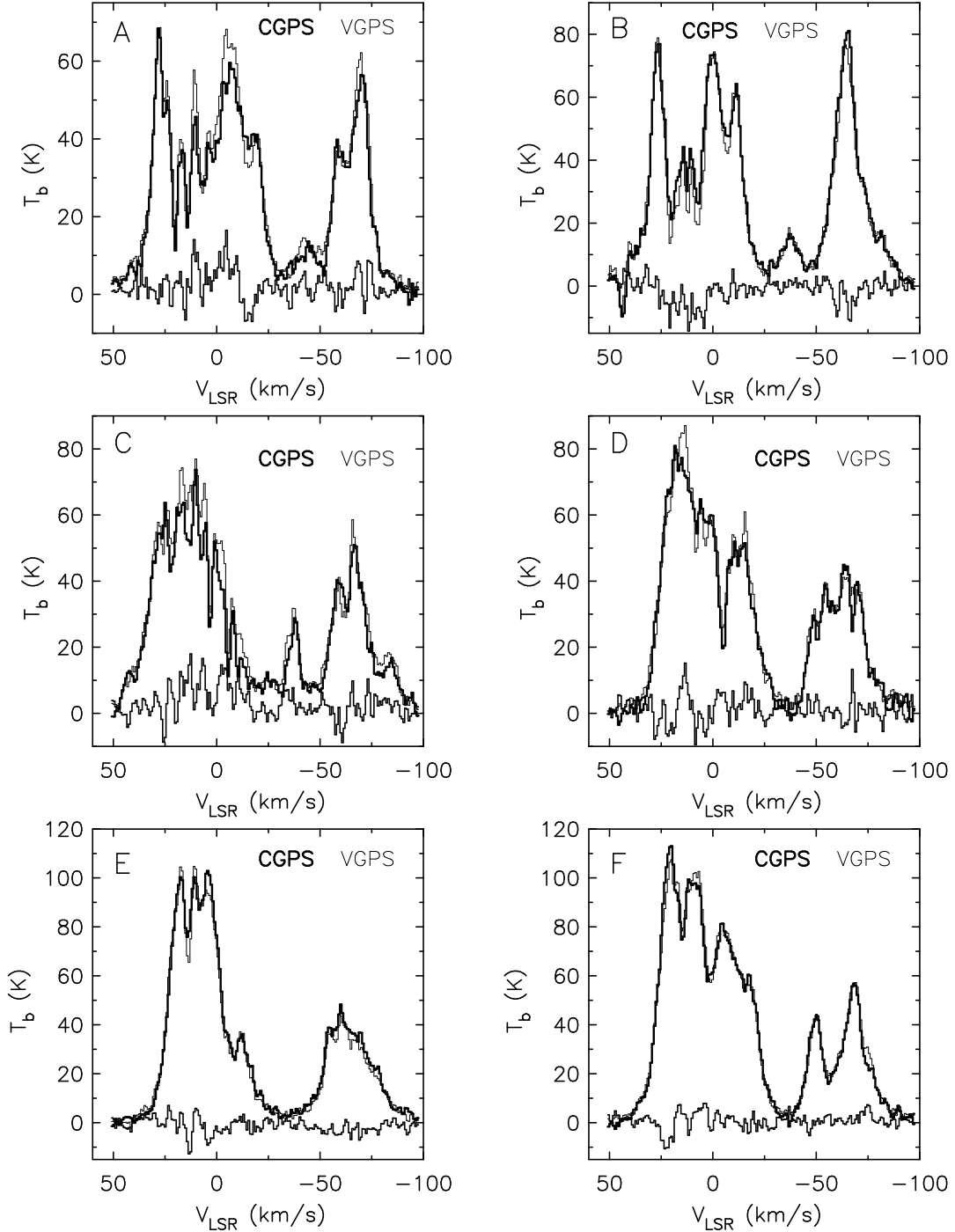}}
\caption{ Comparison of VGPS and CGPS \HI\ line profiles. Shown are
single-position profiles from the CGPS (thick histogram), the VGPS
(thin histogram), and the difference (VGPS$-$CGPS) for six locations.
Panels A-D show lines of sight with significant absorption of
continuum emission. Panels E and F show profiles which are
representative for the general field. The theoretical rms of the
difference profile is $4.4\ \rm K$. The surveys show very good
agreement, but some systematic differences on the level of 5 to 10 K
are found to persist over several channels. Possible origins of these
differences are discussed in Section~\ref{VGPSCGPS-sec}.  Locations of
the profiles in ($l$,$b$) are A: ($64\fdg13$,$-0\fdg47$), B:
($63\fdg16$,$+0.45$), C: ($63\fdg00$,$+0.81$), D:
($65\fdg31$,$-0\fdg22$), E: ($66\fdg24$,$-0\fdg31$), F:
($64\fdg94$,$-0\fdg27$).
\label{VGPSCGPS-profile}
}  
\end{figure}

Figure~\ref{VGPSCGPS-profile} compares \HI\ line profiles from the
CGPS and the VGPS. The CGPS and VGPS \HI\ emission profiles agree very
well, but comparing \HI\ emission profiles is not a sensitive test of
the consistency for small-scale structure, because the signal is
dominated by the single-dish surveys.  Line profiles affected by
continuum absorption are an exception, because the depth of the
absorption profile is very sensitive to
resolution. Figure~\ref{VGPSCGPS-profile} therefore focuses on
continuum absorption profiles.

Judging the significance of the differences between the two datasets
involves the noise in the images at the resolution of $2' \times 1'$.
The rms noise in channels free of line emission in the overlap region,
avoiding noisy edges of the CGPS mosaic, is 1.8 K for the CGPS,
and 1.2 K for the VGPS. These noise amplitudes are consistent with the
theoretical noise levels in DRAO spectral line images at the
declination of the overlap region \citep{landecker2000}, and for VGPS
data convolved to the $2' \times 1'$ resolution of the DRAO images.
The theoretical off-line noise in the difference profile is therefore
2.2 K (rms). In channels where the \HI\ brightness is high ($\sim$100
K), the noise is approximately twice the off-line noise (see
Section~\ref{calibration-sec}).  Differences between the \HI\ profiles
should therefore be compared to an expected rms noise amplitude 4.4 K.
The measured rms amplitude of the difference image per channel is
$2.2\ \rm K$ per velocity channel in the absence of line emission (as
expected), and $5.8\ \rm K$ in channels where the \HI\ emission is
brightest.  The rms amplitude of the difference image over the overlap
area is $\sim 30\%$ higher than the theoretical noise amplitude of the
difference image.  Locally, the differences in
Figure~\ref{VGPSCGPS-profile} are usually less than 10 K, but
occasional differences as large as $\pm 20\ \rm K$ (4 to 5$\sigma$)
exist.

Figure~\ref{VGPSCGPS-diff} shows the difference VGPS$-$CGPS in a
single velocity channel at $+9\ \kms$.  Inspection of the difference
images shows that residuals on small angular scales usually coincide
with emission features in the original images. Some ripple-like
artifacts in the highest rows of VGPS fields ($b = \pm 2\degr$) are
the result of imperfect convergence of the maximum entropy
deconvolution (e.g., Figure~\ref{VGPSCGPS-diff} near $l=66\fdg25$,
$b=-2\degr$). Differences associated with mainly small-scale \HI\
emission are also believed to originate from the deconvolution.  In
rare cases these differences appear as a measurable velocity
difference between features seen in the CGPS and in the VGPS. However,
an analysis of narrow continuum absorption features
(Figure~\ref{VGPSCGPS-profile}) showed that in general the velocities
in the surveys agree very well.

However, the most conspicuous structure in the difference image is a
wave pattern with a wavelength of order $1\fdg5$ which does not
resemble any structure in the \HI\ emission. The pattern is seen with
the same phase and orientation in channels where the average
brightness of the \HI\ line is high, but its amplitude varies with the
brightness of the \HI\ line. The pattern may be isolated by filtering
in the Fourier domain. Subtracting the wave pattern in
Figure~\ref{VGPSCGPS-profile} decreases the rms residuals to $5.2\ \rm
K$.  This is more than the expected value of 4.4 K. If the remaining
difference is entirely the result of the image deconvolution, then the
inconsistencies because of the image deconvolution are less than 3 K
(rms).  The crests of the wave pattern are aligned with the field
centers of the CGPS or possibly the direction of right ascension and
the wavelength corresponds within the errors with the separation of
the CGPS fields. The wavelength of the pattern is more than three
times the field separation of the VGPS. This excludes the VLA data as
a possible origin. Also, the zero-spacing datasets of the CGPS and
VGPS were taken with a primary beam of $30'$ and $9'$, excluding a
possible problem with the primary beam of these telescopes as the
origin.  Both zerospacing datasets were observed scanning across the
Galactic plane at constant longitude. It is not likely this observing
pattern would produce a wave pattern with the orientation seen in
Figure~\ref{VGPSCGPS-diff}.  This leaves the DRAO ST as the most
likely origin of the wave pattern.

The sign of the fluctuations in Figure~\ref{VGPSCGPS-diff} suggests
that the CGPS intensity is systematically higher at larger distances
from the field center in the declination direction. This could
indicate that the primary beam of the DRAO ST is slightly elliptical,
with a larger size in the declination direction than predicted by the
axially symmetric model adopted in the mosaicking
\citep{taylor2003}. This hypothesis was confirmed by a reanalysis of
point sources fluxes measured for the CGPS field registration
\citep{taylor2003} provided by S. J. Gibson.  In the field
registration process, fluxes of compact continuum sources are measured
in individual CGPS fields and compared with fluxes listed in the
NVSS. This process has accumulated $\sim 17,000$ flux measurements
over the years of observations for the CGPS. This dataset was analyzed
for systematic changes in the ratio of CGPS flux to NVSS flux as a
function of position angle in the field. It was found that the flux
ratio CGPS/NVSS was on average a few percent higher along the
declination axis. This systematic effect can be removed by applying a
slightly elliptical primary beam model for the DRAO ST, with axial
ratio $0.975 \pm 0.005$. This value is within the uncertainties of a
previous determination of the primary beam shape of individual DRAO
antennas \citep{dougherty1995}, but the comparison with the VGPS shows
that it has a significant effect in wide-field imaging of the Galactic
\HI\ line.

This comparison shows that despite the critical deconvolution step in
the VGPS image processing, the VGPS images are in general very
consistent with CGPS images which did not require image
deconvolution. We estimated that inconsistencies in the convergence of
the VGPS image deconvolution result in differences less than $\sim$3 K
(rms).  We have found evidence for a small ellipticity in the
primary beam of the DRAO ST, which creates a measurable effect in
wide-field images of bright Galactic \HI\ emission. It should be noted
that the overlap area between the CGPS and the VGPS is in a part of
the sky which is most favorable for the VGPS.

\subsection{VGPS and SGPS}

The Southern Galactic Plane Survey (SGPS) covers a large piece of the
Galactic plane in the southern hemisphere between $l = -107\degr$ and
$l = +20\degr$ and $|b| < 1\degr$. The interferometer used is the ATNF
Australia Telescope Compact Array (ATCA). Zero spacing data is
provided by the multi-beam $64\ \rm m$ Parkes telescope. The ATCA is
mainly an east-west array but it has a north-south arm which was used
for the northernmost part of the SGPS (phase II), including the
overlap region with the VGPS, to obtain more complete sampling in the
{\it uv} plane.  The ATCA elements are similar in size to the VLA
antennas, so individual SGPS fields are similar in size to VGPS
fields. Mosaicking speed is improved at the expense of continuity of
antenna tracks in the {\it uv} plane, in particular at longer
baselines. The density of sampling of the {\it uv} plane by the ATCA
decreases faster with baseline length than for the VLA. The longer
baselines of the ATCA are included in SGPS datasets used to study
continuum absorption, but not in the \HI\ spectral line cubes. This
limits the resolution of the SGPS \HI\ spectral line images to $\sim
2'$ (FWHM).  The SGPS and VGPS use the same software for mosaicking
and image deconvolution, but there are some differences. The SGPS data
acquisition and image processing was described by \citet{mcclure2001}
and \citet{mcclure2005}.

The area of overlap between the VGPS and SGPS is smaller than the
overlap between VGPS and CGPS. Also, SGPS images have $3\arcmin
\times 2\arcmin$ resolution in the overlap area, compared to $2\arcmin
\times 1\arcmin$ for the CGPS. The shape of the synthesized beam of
the SGPS changes rapidly across the overlap area. This is relevant for
comparing absorption profiles, because continuum sources and their
associated absorption are not cleaned separately in the SGPS. Absorbed continuum
sources are convolved with the {\it local} dirty beam which may have a
different shape than the {\it global} restoring beam adopted for the
line emission.  The comparison between SGPS and VGPS images presented
here is therefore less detailed. However, this comparison is useful
because it is done in the part of the IGPS survey area which is most
difficult to observe from the point of view of both the VGPS and the
SGPS. It is therefore a comparison of images made under the least
favorable conditions.

\begin{figure}
\includegraphics[angle=0,scale=.8]{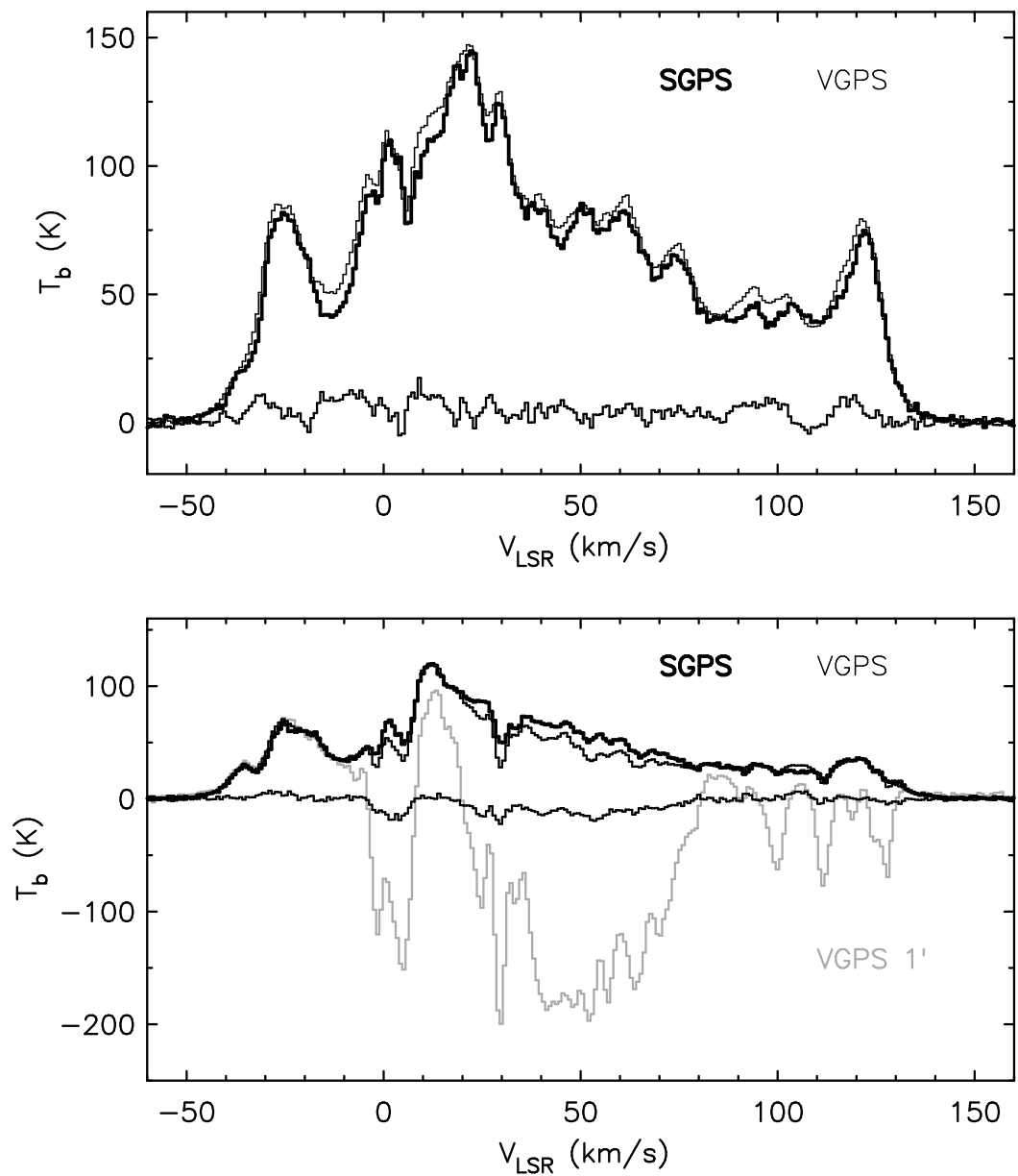}
\caption{ Comparison of VGPS and SGPS \HI\ line profiles. The overlap
region of the VGPS with the SGPS is the most difficult part of the
Galactic plane to observe for both surveys.  Shown are single-position
profiles from the SGPS (thick histogram), the VGPS convolved to the
$3' \times 2'$ beam of the SGPS (thin histogram), and the difference
(VGPS$-$SGPS) for two locations.  The top panel shows a representative
\HI\ emission profile at ($l$,$b$) = ($19\fdg19$,$+0\fdg07$). The
bottom panel shows the strongest absorption profile in the overlap
area at ($l$,$b$) = ($19\fdg61$,$-0\fdg23$). The gray profile in the
bottom panel is the full-resolution ($1'$ FWHM) VGPS absorption
profile at the same location, illustrating that a small discrepancy in
the beam size may contribute significantly to the residuals.
\label{VGPSSGPS-profile}
}  
\end{figure}

Figure~\ref{VGPSSGPS-profile} compares SGPS and VGPS line profiles
with the same spatial resolution at two locations. The upper panel
shows a representative \HI\ emission line profile.  The correspondence
between the VGPS and SGPS profiles is generally as good as between the
CGPS and VGPS. The bottom panel in Figure~\ref{VGPSSGPS-profile} shows
a continuum absorption profile. Here, the quality of the
correspondence between the profiles depends critically on our ability
to match the beam of the SGPS at the location of the source. For
comparison, the full-resolution VGPS profile is also shown.

It is found that the VGPS and SGPS \HI\ images (not shown) are
consistent to a similar degree as the consistency between the CGPS and
VGPS. The comparison between VGPS and SGPS is somewhat limited because
of the smaller area of overlap and the varying beam of the SGPS data.

\section{Panoramic images of the Galactic plane}

The high-resolution images of the VGPS open the opportunity for a
variety of research on the interstellar medium in the first Galactic
quadrant. The images also reveal objects which would not be noticed in
previous surveys with lower resolution. \citet{stil2004} reported
the discovery of a small \HI\ shell in the inner Galaxy. The upper
limit to any thermal emission from this shell derived from the VGPS
continuum image, imposed strong constraints on the interpretation of
this shell as a stellar wind bubble. \citet{lockman2003a} and
\citet{stil2006} reported the discovery of small \HI\ clouds with
large forbidden velocities from the VGPS data. These clouds with
masses of tens of solar masses and radii of a few parsec may be the
disk analogy of similar clouds seen at large distances from the
Galactic plane \citep{lockman2002}. \citet{stil2006} demonstrate the
importance of arcminute-resolution images in the discovery of these
clouds.

The VGPS data are presented in Figure~\ref{mosaics-fig} as panoramic
images of the Galactic plane in 21-cm continuum and \HI\ line
emission.  The continuum images display a strong increase in
complexity of structure as Galactic longitude decreases, reflecting
the higher rate of star formation in the inner Galaxy.  At lower
longitudes, the Galactic plane is traced by a diffuse layer of
emission approximately $1\degr$ thick centered on $b=0\degr$. This
layer may be the thin disk identified at 408 MHz by
\citet{beuermann1985}.  A large number of shells, filaments and
center-filled diffuse emission regions is seen concentrated toward the
Galactic plane, with occasional objects located a degree or more from
$b=0\degr$. Many compact continuum sources are also present. At
longitudes $l \gtrsim 55\degr$, the Galactic plane is much less
discernible in the continuum images. Some small diffuse emission
regions and numerous point sources are seen on top of an extended
otherwise featureless background of continuum emission. The structures
in the VGPS continuum images, with angular scales ranging from several
degrees down to the resolution limit of the VGPS, represent processes
that operate on a wide range in physical size, time scale, and
energy. Many unresolved sources are distant radio galaxies, but
some compact Galactic \HII\ regions and supernova remnants may also
appear unresolved in the VGPS.

The brightest continuum sources in the VGPS are W49 ($l=43\degr$) and
W51 ($l=49\degr$). The VGPS continuum images contain artifacts within
a distance $\sim 1\degr$ around these sources. The dynamic range is
limited because self-calibration for a source far from the field
center is not possible (see Section~\ref{selfcal-sec}). This affects
mainly fields surrounding the bright source.  Sidelobes of a bright
continuum source outside the survey area are present at $l \approx
35\fdg5$, $b \approx -1\fdg3$.

The bottom panels in Figure~\ref{mosaics-fig} show \HI\ emission at
$+3.5\ \kms$. \HI\ at this velocity arises predominantly near the
solar circle (Galactocentric radius 8.5 kpc), and thus both locally
and from gas 6.6 to 16 kpc distant, depending on longitude.  The VGPS
covers only a small fraction of the scale height of local gas, but at
the far side the latitude range $b = \pm 1\fdg3$ spans more than 700
pc (almost twice the FWHM thickness of the \HI\ layer) perpendicular
to the Galactic plane at $l = 20\degr$. Any tendency of \HI\ emission
to be concentrated toward the Galactic plane is therefore most likely
the result of emission on the far side. At longitudes $l\lesssim
30\degr$, \HI\ emission in Figure~\ref{mosaics-fig} is brighter around
$b = 0\degr$ than at $b = \pm 1\fdg3$. Here we see distant \HI\ at the
far side on the solar circle through local \HI.  This distinction
disappears for $l\gtrsim 30\degr$ because the FWHM thickness of the
\HI\ layer on the far side becomes comparable to or larger than the
latitude range covered by the VGPS.

The local \HI\ is widely distributed in latitude, but it is far from
featureless. Most of the structure in the local \HI\ is difficult to
separate from structure in emission from the far side, but there are
exceptions. A band of decreased \HI\ brightness temperature runs
approximately perpendicular to the Galactic plane between $l=20\degr$
and $l=22\degr$.  This dark band is part of a large cloud of cold \HI\
discovered by \citet{heeschen1955} and mapped by \citet{riegel1972}
seen in 'self-absorption' against brighter background \HI\ emission.
\HI\ self-absorption features may represent a transition or a
boundary between molecular and atomic hydrogen.  The high-resolution
VGPS images reveal delicate substructure in this large cloud of cold
hydrogen.  Many \HI\ self-absorption features are seen in VGPS images
at positive velocities on a wide range of angular scales.

Measurements of the opacity of continuum absorption provide an
estimate of the spin temperature of \HI\ and therefore the thermal
structure of the gas. Continuum absorption profiles can also help
resolve the near-far distance ambiguity of kinematic distances for
sources of continuum emission in the inner Galaxy.  Many resolved and
unresolved continuum sources are seen in absorption in the \HI\
images. These appear as a negative imprint of the continuum image onto
the \HI\ images in Figure~\ref{mosaics-fig}. However, this imprint is
not a perfect copy of the continuum image. A good example is the area
between $l=18\degr$ and $l=20\degr$, where some continuum sources are
seen in absorption, whereas other sources of similar brightness are
not. This clearly illustrates the potential of the VGPS data to study
the location and the properties of cold \HI\ clouds in the Galaxy.

Any structural resemblance between \HI\ emission and continuum
emission would be unusual. Continuum emission originates mainly from
ionized gas or relativistic electrons. The 21-cm line of \HI\ traces
neutral atomic gas. Apart from the negative imprint of the continuum
image because of continuum absorption, there is almost no similarity
between the \HI\ and continuum emission in Figure~\ref{mosaics-fig}.
A peculiar exception is a thin filament of continuum emission at
$l=38\fdg35$, $0\degr \le b \le 0\fdg7$. An \HI\ filament with the
same thickness and center line is observed in the VGPS. It is visible
in Figure~\ref{mosaics-fig}, but the velocity of the \HI\ filament is
actually near $0\ \kms$, where it is much more conspicuous against the
other emission.

\section{Conclusions}

\HI\ spectral line and 21-cm continuum images of the Galactic plane
between $l = 18\degr$ and $l = 67\degr$ from the VLA Galactic Plane
Survey (VGPS) are presented. The VGPS data will be made available on
the World Wide Web through the Canadian Astronomy Data Centre (CADC).

The calibration of the VGPS images was made consistent with the
calibration of the NVSS \citep{condon1998}. Initially fluxes of
compact sources in VGPS images were found to be systematically below
the fluxes listed in the NVSS by up to 30\%. A correction to the flux
scale of the VGPS was developed, based on the noise amplitude in the
visibility data. This procedure can be applied to any VGPS field,
including those without sufficient point sources to compare the flux
directly with the NVSS, and those fields which are filled with complex
extended emission. After this correction, the fluxes of compact
continuum sources in the VGPS were found to be consistent with NVSS
fluxes to within 5\%.

VGPS \HI\ images were compared with those from the CGPS and SGPS in
regions of overlap and show good agreement between the three surveys,
although the rms amplitude of the difference image (VGPS$-$CGPS) is
approximately 30\% higher than the fluctuations expected from the
theoretical noise levels alone.  Small systematic differences between
the VGPS and CGPS originate from imperfections of the image
deconvolution applied in the VGPS image processing, and from a wave
pattern which resembles the effect of ellipticity in the primary beam
of the DRAO ST.

\section*{Acknowledgements}
The National Radio Astronomy Observatory is a facility of the National
Science Foundation operated under cooperative agreement by Associated
Universities, Inc.  The VGPS is supported by a grant to A.R.T. from
the Natural Sciences and Engineering Council of Canada.  This research
was supported in part by the National Science Foundation through
grants AST 97-32695 and AST 03-07603 to the University of Minnesota.
JMS thanks Richard Gooch for adding code to the KARMA software which
reads miriad {\it uv} datasets used in our automated flagging program,
and Steven J. Gibson for providing the CGPS field registration
data. The authors thank the anonymous referee for useful comments on
the manuscript.

{}

\begin{deluxetable}{lr} 
\tablecolumns{2}
\tablewidth{0pc} 
\tablecaption{VGPS parameters} 
\tablehead{ 
\colhead{Quantity } &  Value \ \ \ \  
}   
\startdata
Survey area            &  $|b| < 1\fdg3$ \ \ \ \ $18\degr < l < 46\degr$ \\
                       &  $|b| < 1\fdg9$ \ \ \ \ $46\degr < l < 59\degr$ \\ 
                       &  $|b| < 2\fdg3$ \ \ \ \ $59\degr < l < 67\degr$ \\ 
Angular resolution (FWHM)    &  $60\arcsec \times 60\arcsec$   \\
Spectral resolution (FWHM)   &  $1.56\ \kms$  \\
Channel width          &   $0.824\ \kms$ \\
Noise in continuum\tablenotemark{a}     &   $0.3\ \rm K$ \\
Noise per channel\tablenotemark{a}      &  $2\ \rm K$ \\
$T_b/S_\nu$                             &  168 K/Jy \\
\enddata
\tablenotetext{a}{Noise levels may be different from these representative values
depending on location and velocity (\HI\ line).}
\label{VGPSpar-tab}
\end{deluxetable}

\begin{deluxetable}{lrrrrrr} 
\tablecolumns{7}
\tablewidth{0pc} 
\tablecaption{Fits of Equation (2)} 
\tablehead{ 
\colhead{Date} &  \colhead{$a_L$}  &  \colhead{$b_L$}  & \colhead{$\sigma_L$} &  \colhead{$a_R$}  &  \colhead{$b_R$}  & \colhead{$\sigma_R$} 
}   
\startdata
07/23 &    1.26 &   1.033 &   0.088 &     $\ldots$ &  $\ldots$  &  $\ldots$ \\ 
07/25 &    1.10 &   1.078 &   0.061 &     1.14 &   1.169 &    0.057 \\ 
07/28 &    1.28 &   1.079 &   0.060 &     1.32 &   1.182 &    0.061 \\ 
07/29 &    1.52 &   1.075 &   0.094 &     1.54 &   1.216 &    0.093 \\ 
08/02 &    1.37 &   1.021 &   0.052 &     1.41 &   1.135 &    0.049 \\ 
08/04 &    1.38 &   1.133 &   0.055 &     1.44 &   1.276 &    0.051 \\ 
08/05 &    1.48 &   1.090 &   0.031 &     1.58 &   1.187 &    0.037 \\ 
08/08 &    1.50 &   1.056 &   0.058 &     1.42 &   1.215 &    0.051 \\ 
08/10 &    1.41 &   1.079 &   0.076 &     1.51 &   1.166 &    0.071 \\ 
08/14 &    1.31 &   1.121 &   0.061 &     1.31 &   1.246 &    0.059 \\ 
08/17 &    1.46 &   1.102 &   0.063 &     1.55 &   1.231 &    0.066 \\ 
08/19 &    1.30 &   1.055 &   0.026 &     1.31 &   1.141 &    0.029 \\ 
08/22 &    1.26 &   1.166 &   0.055 &     1.28 &   1.228 &    0.054 \\ 
08/24 &    1.42 &   1.110 &   0.051 &     1.45 &   1.272 &    0.061 \\ 
08/29 &    0.89 &   1.118 &   0.090 &     0.91 &   1.228 &    0.093 \\ 
08/31 &    1.41 &   1.115 &   0.039 &     1.51 &   1.262 &    0.040 \\ 
09/05 &    1.40 &   1.143 &   0.042 &     1.54 &   1.287 &    0.046 \\ 
09/07 &    1.64 &   1.115 &   0.058 &     1.75 &   1.260 &    0.058 \\ 
09/11 &    1.39 &   1.132 &   0.034 &     1.54 &   1.272 &    0.034 \\ 
09/14 &    1.17 &   0.933 &   0.097 &     1.18 &   1.243 &    0.109 \\ 
09/15 &    1.41 &   1.147 &   0.066 &     1.42 &   1.314 &    0.061 \\ 
09/17 &    1.29 &   1.011 &   0.041 &     1.40 &   1.127 &    0.041 \\ 
09/18 &    1.33 &   1.258 &   0.163 &     1.79 &   1.351 &    0.174 \\ 
09/19 &    1.32 &   1.139 &   0.065 &     1.46 &   1.284 &    0.061 \\ 
09/21 &    1.15 &   1.237 &   0.088 &     1.26 &   1.398 &    0.101 \\ 
09/24 &    1.15 &   0.996 &   0.044 &     1.12 &   1.083 &    0.055 \\ 
09/29 &    1.34 &   1.146 &   0.054 &     1.51 &   1.233 &    0.050 \\ 
09/30 &    1.32 &   1.173 &   0.060 &     1.32 &   1.273 &    0.065 \\ 
\enddata
\label{elevationfit-tab}
\end{deluxetable}

\begin{figure}
\caption{ {\bf (Provided as separate gif image)} The Sensitivity
Function of the VGPS.  A representative section of the primary survey
area is shown.  Contours are drawn at the 50\%, 80\%, 90\%, 95\%, and
98\% levels of the maximum sensitivity in the mosaic.  This function
reflects the Gaussian primary beam shape of the VLA and the grid of
pointing centers indicated by the crosses.  The latitude coverage of
the survey widens by two more rows of pointing centers for longitudes
greater than $45\fdg7$, and a further two rows for $l > 59\fdg7$.
\label{jd1}
}  
\end{figure}

\begin{figure}
\caption{ {\bf (Provided as separate gif image)} Band averaged
ampscalar as a function of position in the sky. Top panel: smallest
ampscalar for each field in gray scales. Middle panel: smallest
ampscalar per field, corrected for elevation-dependent spillover to
the ground. Bottom panel: VGPS continuum image. The predominance of
the observing pattern of blocks of 6 fields in the upper panel shows
that spillover to the ground is important compared with Galactic
emission almost everywhere in the VGPS. When the elevation-dependent
contribution is eliminated, as described in the text, a clear
correlation with bright Galactic emission is seen.
\label{Ecor}
}  
\end{figure}

\begin{figure}
\caption{ {\bf (Provided as separate gif image)} Sampling of the {\it
uv} plane and the synthesized beam for a representative field in the
center of the survey area. This field is centered on
$(l,b)=(42\fdg5,-0\fdg1)$. It was observed three times at hour angles
$-2\fh7$, $-0\fh2$, and $+2\fh7$.  Left: sampling function of the {\it
uv} plane. Right: the corresponding synthesized beam, with peak
amplitude equal to 1. Gray scales increase linearly from $-0.05$ to
$0.30$ in steps of $0.05$.
\label{snapshotUV}
}  
\end{figure}

\begin{figure}
\caption{ {\bf (Provided as separate gif image)} Difference image
(VGPS$-$CGPS) in the overlap area of VGPS and CGPS phase II for a
single velocity channel at $+9\ \kms$. Black areas on the right are
outside the CGPS phase II survey area. The gray scale is linear from
$-10\ \rm K$ to $+10\ \rm K$. Crosses mark the CGPS field centers. The
directions of increasing right ascension and declination are indicated
by arrows. The pattern of diagonal dark and light bands aligned with
the CGPS field centers originates from the DRAO synthesis
telescope. Some artifacts in the VGPS are visible at latitudes
$\pm2\degr$.
\label{VGPSCGPS-diff}
}  
\end{figure}

\begin{figure}
\caption{{{\bf (Provided as separate gif image)} VGPS
panoramic image of the Galactic plane.  Top panel: Continuum
image. Gray scales are logarithmic from 0 to 90~K.  Bottom: \HI\
spectral line image at velocity $+3.5\ \kms$.  Gray scales are linear
from 0 to 125~K. The extent of the spectral line images is limited by
the coverage of the GBT data. The continuum images display the
Effelsberg data only outside the area covered by the VLA data (see
Table~\ref{VGPSpar-tab}}).
\label{mosaics-fig}
}  
\end{figure}

\addtocounter{figure}{-1}
\begin{figure}
\caption{ {\bf (Provided as separate gif image)} Continued.
}  
\end{figure}

\addtocounter{figure}{-1}
\begin{figure}
\caption{ {\bf (Provided as separate gif image)} Continued.
}  
\end{figure}

\addtocounter{figure}{-1}
\begin{figure}
\caption{ {\bf (Provided as separate gif image)} Continued.
}  
\end{figure}


\begin{thebibliography}{}

\bibitem[Beuermann et~al.(1985)]{beuermann1985} Beuermann, K., Kanbach, G., and Berkhuijsen, E. M. 1985, \aap, 153, 17

\bibitem[Condon et~al.(1998)]{condon1998} Condon, J. J., Cotton, W. D., Greisen, E. W., Yin, Q. F., Perley, R. A., Taylor, G. B., \& Broderick, J. J. 1998, \aj, 115, 1693

\bibitem[Cornwell \& Evans(1985)]{cornwell1985} Cornwell, T. J., \& Evans, K. F. 1985, \aap, 143, 77

\bibitem[Dickey \& Lockman(1990)]{DL1990} Dickey, J. M., \& Lockman, F. J. 1990, \araa, 28, 215

\bibitem[Dougherty(1995)]{dougherty1995} Dougherty, S. M. 1995, DRAO technical report, October 1995

\bibitem[Heeschen(1955)]{heeschen1955} Heeschen, D. S. 1955, \apj, 121, 569

\bibitem[Higgs \& Tapping(2000)]{higgs2000} Higgs, L. A., \& Tapping, K. F. 2000, \aj, 120, 2471

\bibitem[H\"ogbom(1974)]{hogbom1974} Hogbom, J. A. 1974, A\&AS, 15, 417

\bibitem[Landecker et~al.(2000)]{landecker2000} Landecker, T. l., Dewdney, P. E., Burgess, T. A., Gray, A. D., Higgs, L. A., Hoffmann, A. P., Hovey, G. J., Karpa, D. R., Lacey, J. D., Prowse, N., Purton. C. R., Roger R. S., Willis, A. G., Wyslouzil, W., Routledge, D., \& Vaneldik, J. F. 2000, A\&AS, 145, 509

\bibitem[Lockman(2002)]{lockman2002} Lockman, F. J. 2002, \apj,  580, L47

\bibitem[Lockman \& Stil(2003)]{lockman2003a} Lockman, F. J., \& Stil, J. M. 2004, in ASP Conf. Ser. 317, Milky Way Surveys: The Structure and Evolution of Our Galaxy, ed. D. Clemens, R. Y. Shah, \& T. Brainerd, (San Francisco: ASP), 20

\bibitem[McClure-Griffiths et~al.(2001)]{mcclure2001} McClure-Griffiths, N. M., Green A. J., Dickey J. M., Gaensler, B. M., Haynes, R. F., \& Wieringa, M. H. 2001, \apj, 551, 394 

\bibitem[McClure-Griffiths et~al.(2005)]{mcclure2005} McClure-Griffiths, N. M., Dickey, J. M., Gaensler, B. M., Green, A. J., Haverkorn, M., \& Strasser, S. 2005, \apjs, 158, 178

\bibitem[Reich \& Reich(1986)]{reich1986} Reich, P.,\& Reich, W. 1986, A\&AS, 63, 205

\bibitem[Reich et~al.(1990)]{reich1990} Reich, W., Reich, P., \& F\"urst 1990, A\&AS, 83, 539

\bibitem[Riegel \& Crutcher(1972)]{riegel1972} Riegel, K. W., \& Crutcher, R. M. 1972, \aap, 18, 55

\bibitem[Sault et~al.(1996)]{sault1996} Sault, R. J., Staveley-Smith, L., \& Brouw, W. N. 1996, A\&AS, 120, 375    

\bibitem[Stil et~al.(2004)]{stil2004} Stil, J. M., Taylor, A. R., Martin, P. G., Rothwell, T. A., Dickey, J. M., McClure-Griffiths, N. M. 2004, \apj, 608, 297

\bibitem[Stil et~al.(2006)]{stil2006} Stil, J. M., Lockman, F.J., Taylor, A. R., Dickey, J. M., Kavars, D. W., Martin, P. G., Rothwell, T. A., Boothroyd, A. I., McClure-Griffiths, N. M. 2006, \apj, 637, 366

\bibitem[Taylor et~al.(2002)]{taylor2002} Taylor, A. R., Stil, J. M., Dickey, J. M., McClure-Griffiths, N. M., Martin, P. G., Rothwell, T., Lockman, F. J. 2002, in ASP Conf. Ser. 276, Seeing Through The Dust: The Detection Of \HI And The Exploration Of The ISM In Galaxies, ed. A. R. Taylor, T. L. Landecker, \& A. G. Wills (San Francisco: ASP), 68

\bibitem[Taylor et al.(2003)]{taylor2003} Taylor, A. R., Gibson, S. J., Peracaula, M., Martin, P. G., Landecker, T. L., Brunt, C. M., Dewdney, P. E., Dougherty, S. M., Gray, A. D., Higgs, L. A., Kerton, C. R., Knee, L. B. G., Kothes, R., Purton, C. R., Uyaniker, B., Wallace, B. J., Willis, A. G., \& Durand, D. 2003, \aj, 125, 3145 


\bibitem[Taylor, Ulvestad \& Perley(2003)]{TUP2003} Taylor, G. B., Ulvestad, J. S., \& Perley, R. A. 2003, on-line document: The Very Large Array Observational Status Summary, 26 May 2003, {\tt http://www.vla.nrao.edu/astro/guides/vlas/current/vlas.html} 

\bibitem[Williams(1973)]{williams1973} Williams, D. R. W. 1973, \aaps, 8, 505

\end{thebibliography}
\end{document}